\documentclass[lettersize,journal]{IEEEtran}
\usepackage{amsmath,amsfonts}
\usepackage{algorithmic}
\usepackage{algorithm}
\usepackage{array}
\usepackage[caption=false,font=normalsize,labelfont=sf,textfont=sf]{subfig}
\usepackage{textcomp}
\usepackage{stfloats}
\usepackage{url}
\usepackage{verbatim}
\usepackage{graphicx}
\usepackage{cite}

\usepackage{amssymb}
\usepackage{xcolor}
\usepackage{bm}
\usepackage{latexsym}

\def\qed{\hfill $\Box$}
\newtheorem{theorem}{Theorem}
\newtheorem{prop}{Proposition}
\newtheorem{lemma}{Lemma}
\begin{document}

\title{Ordinary Differential Equation-based\\ MIMO Signal Detection}

\author{Ayano Nakai-Kasai,~\IEEEmembership{Member,~IEEE,} 
        \thanks{A. Nakai-Kasai and T. Wadayama are with Nagoya Institute of Technology, Gokiso, Nagoya, Aichi 466-8555, Japan,}
      and Tadashi Wadayama,~\IEEEmembership{Member,~IEEE} 
\thanks{Part of this research was presented at the IEEE Global Communications Conference 2022 (GLOBECOM2022) \cite{globecom}.}
}

\markboth{Journal of \LaTeX\ Class Files,~Vol.~14, No.~8, August~2021}%
{Shell \MakeLowercase{\textit{et al.}}: A Sample Article Using IEEEtran.cls for IEEE Journals}


\maketitle

\begin{abstract}
    The required signal processing rate in future wireless communication systems 
    exceeds the performance of the latest electronics-based processors.
    Introduction of analog optical computation is one promising direction 
    for energy-efficient processing.
    This paper considers a continuous-time minimum mean squared error detection 
    for multiple-input multiple-output systems 
    to realize signal detection using analog optical devices.
    The proposed method is formulated by an ordinary differential equation (ODE) 
    and its performance at any continuous time can be theoretically analyzed.
    Deriving and analyzing the continuous-time system is a meaningful step 
    to verifying the feasibility of analog-domain signal processing in the future systems.
    In addition, considering such an ODE brings byproducts to discrete-time detection algorithms, 
    which can be a novel methodology of algorithm construction and analysis.
\end{abstract}


\section{Introduction}
\label{sec:intro}
\IEEEPARstart{I}{n} future wireless communication systems, beyond 5G and 6G, 
traffic and computational loads at base stations 
for achieving massive connectivity are becoming heavier \cite{6G}.
In particular, 
signal detection methods expected as the key technology in the next generation
will require the order of hundreds of Tera or tens of Peta multiply-accumulate operations per second \cite{Salmani}.
This exceeds the performance of the latest single digital-electronics-based processor depending on graphics processing unit (GPU) and 
the use of multiple processors causes tremendous power consumption \cite{Stroev}.
The progress of computational hardware following Moore's law is reaching a saturation point \cite{Moore}.
Moreover, there is also a computer system throughput limitation caused by data transmission between processor and memory, 
which is known as von Neumann bottleneck \cite{Neumann}.
These limitations hinder the realization of next-generation communication systems.
The use of task-specific hardware \cite{Chetan} may be a potential solution 
but it cannot circumvent the above hardware limitations 
and the large-scale circuits result in significant energy consumption.
Therefore, the development of alternative signal 
processing schemes is necessary to achieve a reasonable signal detection performance with
high energy efficiency.

Analog photonic or optical computing \cite{Stroev,Wu,photonics} is a promising candidate 
for the alternative signal processing scheme that meets the requirements of the next-generation communication systems \cite{Salmani,Yuan}.
Photonic/optical computing, in contrast to electronic computing, 
can receive the benefits of light, 
such as high speed, broader bandwidth, and low heat production. 
{{Integration technology of photonic/optical devices onto silicon chips, known as silicon photonics, 
allows compact, low-cost, and energy-efficient processing, 
and is facilitating photonic/optical computing and the combination with electronic device-based calculation \cite{silicon1,silicon2}.}}
The use of these technologies is already being considered 
in the fields of artificial intelligence that requires training with large amounts of data \cite{silicon1,Wetzstein,Lin,Haensch}, 
and large-scale problems such as combinatorial optimization \cite{Prabhu2} or probabilistic graphical models \cite{Ab}.
Optical computing approach for signal processing in wireless communication is also beginning to be considered.
Salmani et al. proposed a preprocessing method for photonic computing to deal with complex-valued signals in wireless communications \cite{Salmani}.
A recent study by Zhang et al. \cite{HZhang} reported a complex-valued neural network on a photonic chip, 
which motivates the realization of complex-valued signal processing with photonic devices.

Application of optical computing to signal detection 
may be a key technology for overcoming computational bottleneck at base stations.
Typical signal detection methods for massive multiple-input multiple-output (MIMO) systems, 
such as zero-forcing (ZF), linear minimum mean squared error (MMSE) \cite{mimobook}, 
and orthogonal approximate message passing (OAMP) \cite{oamp}, depend on the inverse calculation of a large-scale channel state matrix.
In addition, the inverse calculation may have to be repeated frequently in response to changes in the communication environment \cite{Prabhu}.
Significant signal detection loads at base stations due to these factors 
have become a major bottleneck in the implementation of next-generation systems \cite{6G}.
A promising solution is to employ matrix multiplication as an alternative to inverse calculation 
and use optical computing devices for the multiplication \cite{Wu,Stroev}.
Inverse calculation and multiplication can be in the same order of complexity in electronic computing, 
but multiplication is preferred from a hardware perspective 
{{because the parallelism of optics allows matrix-vector multiplication with $\mathcal{O}(1)$ time complexity \cite{Athale}.}}
For example, programmable Mach-Zehnder interferometers (MZIs) and diffractive optical elements 
provide matrix multiplication with low energy costs in artificial intelligence or combinatorial optimization tasks \cite{Wu,mzi,Zhang,Lin}.
Prabhu et al. \cite{Prabhu2} experimentally demonstrated a system including MZIs for matrix product calculation 
and suggested that computational time and energy can be tremendously reduced.
With these developments in photonic/optical computing 
and attempts to handle complex-valued signals on photonic/optical devices \cite{Salmani,HZhang}, 
signal processing in base stations can be replaced with optical ones in the near future.

\begin{figure}[tb]
    \centerline{\includegraphics[width=\columnwidth]{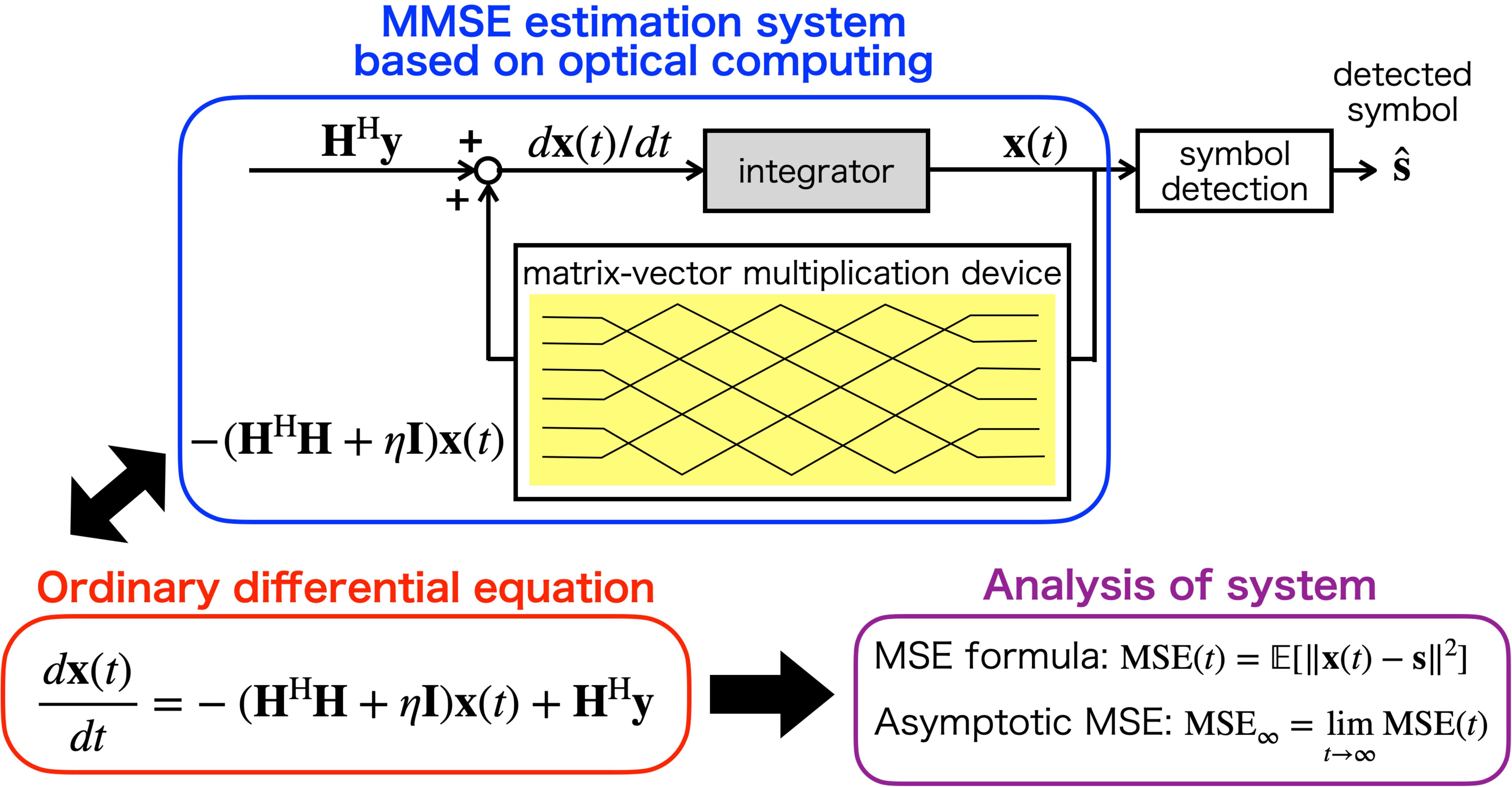}}
    \caption{The main contents of this paper and their relation.
        Implementation of MMSE estimation system can be realized on the basis of optical computing.
        The MSE behavior of the system can be analyzed by considering convergence properties of the ODE.}
    \label{fig:analog}
\end{figure}

In this paper, we consider a signal detection scheme 
based on matrix-vector multiplications 
composed of such analog optical computing devices, 
as a tool to overcome the computing bottleneck at base stations.
As the first step in such an approach, 
this paper deals with {{classical linear}} MMSE detection.
Figure~\ref{fig:analog} shows the main contents of this paper and their relation.
We present a continuous-time MMSE estimation system 
that can potentially be implemented using analog optical computing devices.
The proposed method is derived from \emph{gradient flow dynamics} \cite{Helmke} based on the least square objective function, 
{{where this approach can be applied to many signal processing schemes such as low-density parity-check codes (LDPC) decoding \cite{LDPC} and sparse signal recovery \cite{wadayama2}.}}
The behavior of the system is described by a form of ordinary differential equation (ODE).
In addition, by utilizing the ODE representation, 
mean squared error (MSE) performance of 
the ODE-based MMSE estimation can be theoretically analyzed.

To the best of our knowledge, 
there are no relevant proposals or analyses 
in the previous literature.
Analog optical computing for signal processing 
is still a developing technology from a hardware perspective.
However, analyzing the proposed method 
can be a meaningful step toward 
advancing energy-efficient signal processing using optical devices for wireless communications in the future.

The main contributions of this paper are as follows.
\begin{enumerate}
    \item We propose a continuous-time MMSE estimation method for MIMO systems 
            derived from gradient flow dynamics based on the regularized least square objective function.
            The method includes a regularization parameter that controls the convergence behavior of the estimation method.
    \item An analytical formula of MSE, 
            which is the principal performance measure of signal estimation, 
            is derived in a closed form.  
            From the MSE formula, we immediately derive the asymptotic MSE.
            These analyses enable us to track the quality of the estimation at any time instant.
    \item We introduce a time-dependent regularization parameter for the proposed continuous-time MMSE estimation method 
            to improve convergence performance. 
            An analytical MSE formula can be derived also for the time-dependent system. 
\end{enumerate}

In addition, 
as byproducts of considering such continuous-time MMSE estimation, 
we present discrete-time detection algorithms 
derived by discretizing ODE with numerical solvers \cite{numericalsolutionbook,Eftekhari}.
The discretization considering numerical stability 
and flexible selection of the step size of numerical solver 
leads to higher performance than conventional methods \cite{Dai,jsor}.
This relationship can be observed in the recently proposed neural ODE \cite{RTQChen}, 
which is an ODE that includes a neural network.
The MSE analyses obtained from the continuous-time method 
can be applied to this discrete-time algorithm, 
making the behavior of the algorithm fully traceable.
This analysis does not require any assumption on the system 
although assumptions such as a large system limit are required in some detection algorithms \cite{Hayakawa}.
The benefit indicates the potential for continuous-time signal processing 
as a novel construction and analysis methodology for discrete-time algorithms.

The remainder of this paper is organized as follows.
We introduce the mathematical notations and system model in Section~\ref{sec:pre}.
Sections~\ref{sec:odemmse}--\ref{sec:sim} address continuous-time estimation 
assuming the use of optical computing devices.
The estimation method is proposed in Section~\ref{sec:odemmse}.
Some theoretical properties of this method can be derived, as discussed in Section~\ref{sec:analysis}.
In Section~\ref{sec:odetime}, we show the improvement of the proposed method 
by introducing a time-dependent regularizer.
Verification of the analyses and the comparison of the proposed methods are numerically evaluated in Sect.~\ref{sec:sim}.
Section~\ref{sec:discrete} presents the byproducts of considering ODE on discrete-time signal detection algorithms.
Finally, Section~\ref{sec:conc} concludes the paper.
{{The proposed estimation method was first introduced in a conference paper \cite{globecom}.
The present manuscript additionally introduces the implementation of the proposed method, 
adds detailed numerical evaluations on practical modulations and system models, 
and provides developments to discrete-time algorithms.}}

\section{Preliminaries}
\label{sec:pre}
\subsection{Notation}
In the rest of the paper, we use the following notation.
Superscript $(\cdot)^{\mathrm{H}}$ denotes the Hermitian transpose.
The zero vector and identity matrix are represented by $\bm{0}$ and $\bm{I}$, respectively.
The Euclidean ($\ell_2$) norm is $\|\cdot\|$.
The complex circularly symmetric Gaussian distribution $\mathcal{CN}(\bm{0},\bm{\Sigma})$ 
has a mean vector $\bm{0}$ and a covariance matrix $\bm{\Sigma}$.
The expectation and trace operators are $\mathbb{E}[\cdot]$ and $\mathrm{Tr}[\cdot]$, respectively.
The diagonal matrix is given by $\mathrm{diag}[\ldots]$ with the diagonal elements shown in square brackets.
The matrix exponential $\exp(\bm{A})$ for a matrix $\bm{A}$ is defined by 
$\exp(\bm{A}):=\sum_{k=0}^\infty \bm{A}^k/(k!)$.
The function $\mathrm{mod}(k,s)$ means the remainder of $k/s$ 
and $T_{s}(z)$ is Chebyshev polynomial of the first kind, 
$T_s(\cos\theta) = \cos{(s\theta)}$.

\subsection{First-order Linear ODE}
\label{sec:solution}
Consider a linear ODE with a constant matrix coefficient: 
\begin{equation}
    \frac{d\bm{y}(t)}{dt}=\bm{A}\bm{y}(t),
    \label{eq:exode}
\end{equation}
where $t\geq0$ and $\bm{A}$ is a matrix that is independent of $\bm{y}(t)$ and $t$.
This ODE can be solved analytically using a matrix exponential \cite{numericalsolutionbook}.
The solution is given by 
\[\bm{y}(t)=\exp(\bm{A}t)\bm{y}(0), \]
where $\bm{y}(0)$ denotes the initial value of $\bm{y}(t)$.

The continuous-time dynamical system in \eqref{eq:exode} is asymptotically stable 
if $\bm{y}(t)$ converges to the origin $\bm{0}$ as $t\to\infty$ for all initial conditions $\bm{y}(0)$ \cite{numericalsolutionbook}.
The system is asymptotically stable if and only if the real parts of all the eigenvalues of the matrix $\bm{A}$ are negative \cite{Arnold}.

\begin{figure}[tb]
    \begin{minipage}[b]{0.47\linewidth}
      \centering
      \includegraphics[width=\columnwidth]{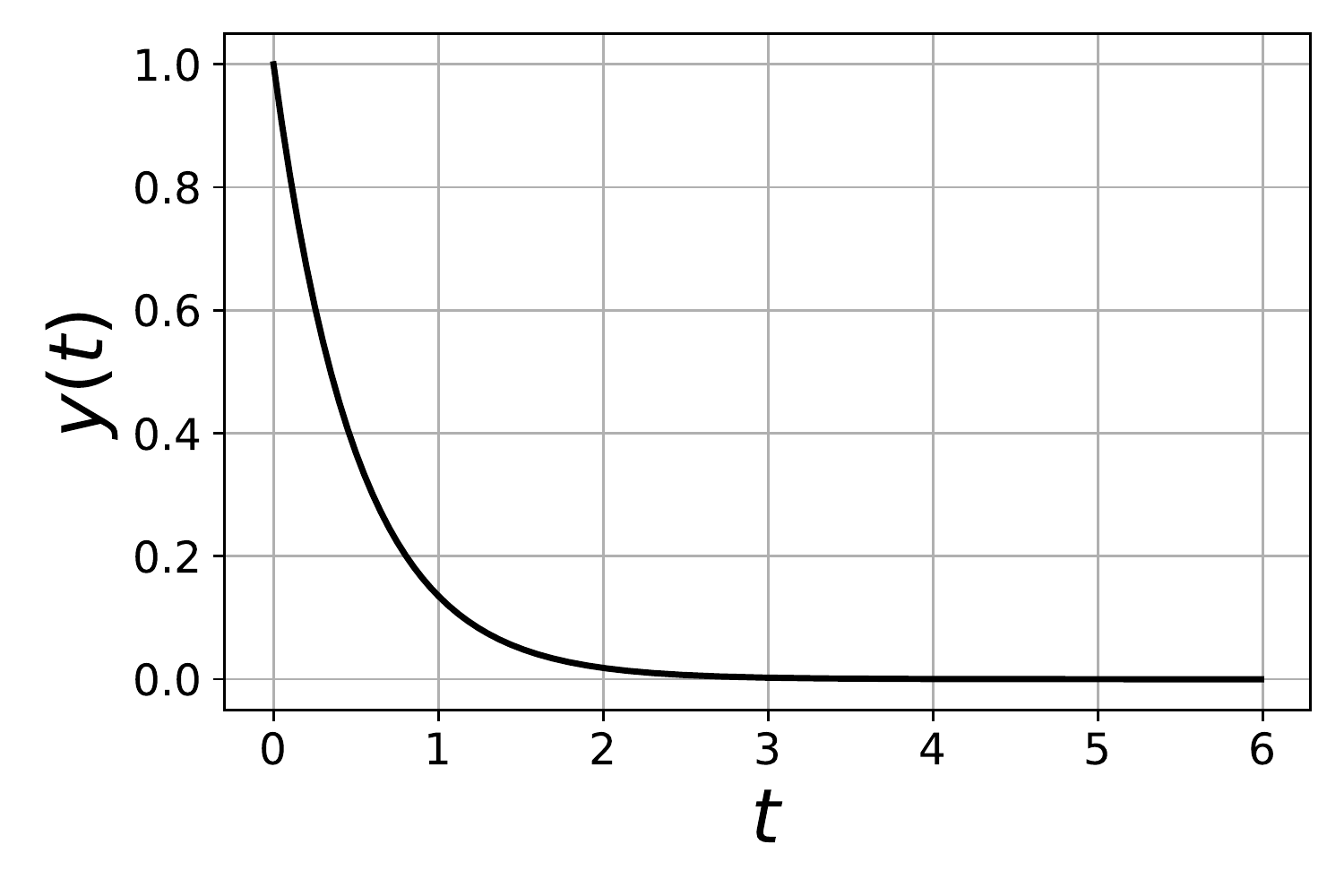}
      \caption{Example of time evolution of ODE $dy(t)/dt=-\lambda y(t)$.}
      \label{fig:ex}
    \end{minipage}
    \hspace{0.04\columnwidth}
    \begin{minipage}[b]{0.47\linewidth}
      \centering
      \includegraphics[width=\columnwidth]{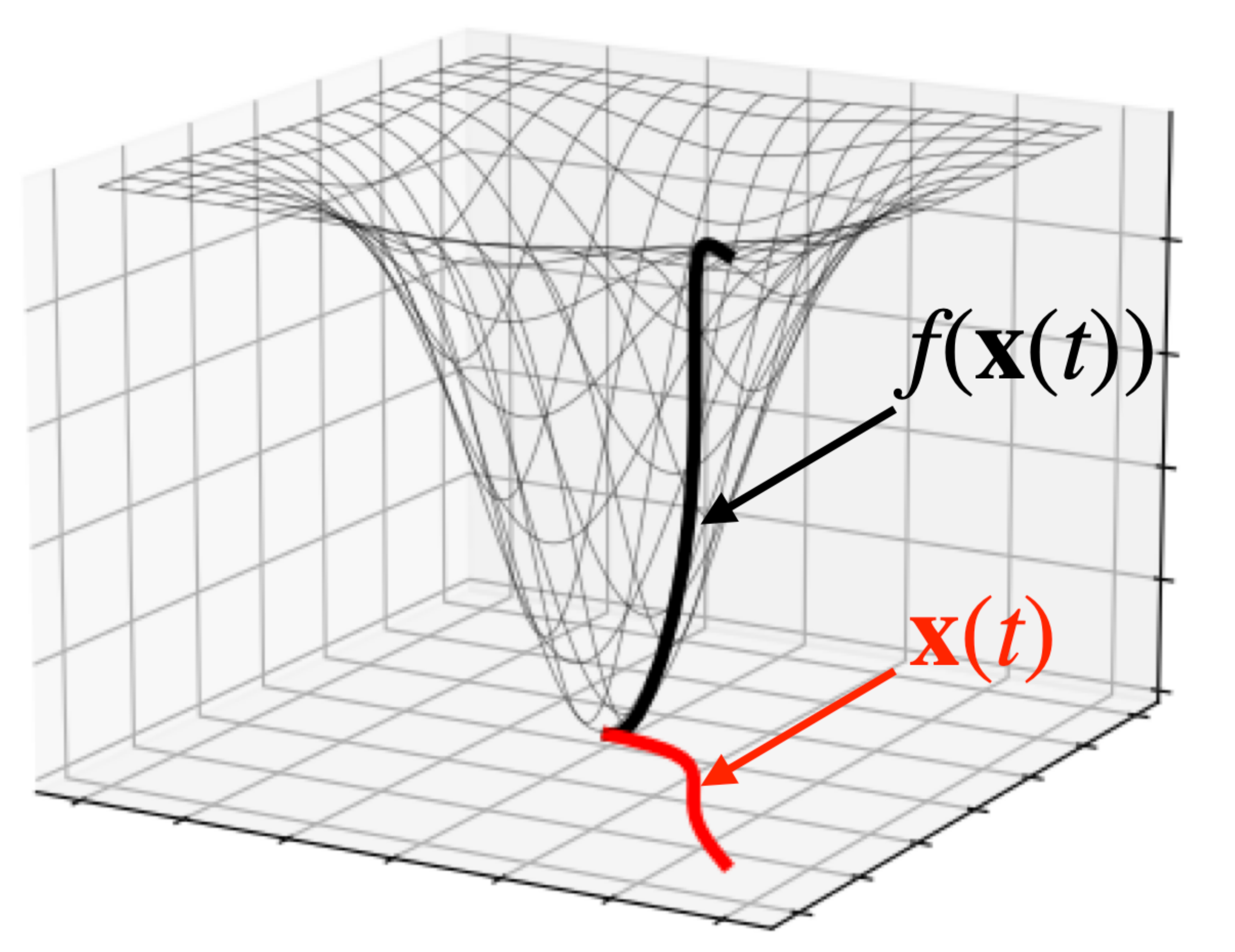}
      \caption{{{Gradient flow dynamics.}}}
      \label{fig:gradientflow}
    \end{minipage}
\end{figure}

A simple example of the linear ODE is $dy(t)/dt=-\lambda y(t)$ $(t\geq0)$, 
where $\lambda\in\mathbb{R}$ is a positive constant.
The solution for the equation is $y(t)=\exp{(-\lambda t)}y(0)$. 
An example of time evolution of the linear ODE when $(\lambda,y(0))=(2,1)$ is presented in Fig.~\ref{fig:ex}.
This system is asymptotically stable.

\subsection{System Model}
\label{sec:system}
In this paper, we consider the following MIMO channel model:
\begin{equation}
    \bm{y} = \bm{Hs} + \bm{w},
    \label{eq:y}
\end{equation}
where $\bm{y}\in\mathbb{C}^m$ is the received signal, 
$\bm{H}\in\mathbb{C}^{m\times n}$ is the channel matrix, 
$\bm{s}\in\mathbb{C}^n$ is the transmitted signal, 
and $\bm{w}\in\mathbb{C}^m$ is the measurement noise that follows $\mathcal{CN}(\bm{0},\sigma^2\bm{I}$).

A linear estimate $\hat{\bm{x}}:=\bm{Wy}$ of the transmitted signal $\bm{s}$ is characterized by the matrix $\bm{W}\in\mathbb{C}^{n\times m}$, 
which is determined according to each estimation method.
Matrix $\bm{W}$ for the linear MMSE estimation \cite{mimobook} can be obtained by minimizing the MSE 
given by $\mathbb{E}[\|\bm{Wy}-\bm{s}\|^2]$.
The resulting MMSE estimate is derived as 
\begin{equation}
    \hat{\bm{x}} = \left(\bm{H}^\mathrm{H}\bm{H}+\sigma^2\bm{I}\right)^{-1}\bm{H}^\mathrm{H}\bm{y}.
    \label{eq:mmse}
\end{equation}

\section{Continuous-time MMSE Estimation}
\label{sec:odemmse}
In this section, we derive a continuous-time MMSE estimation system configurable with analog optical circuits, 
the time evolution of which is described by ODE.

\subsection{Derivation}
A function 
\begin{equation}
    f(\bm{x}) := \|\bm{y}-\bm{Hx}\|^2 + \eta\|\bm{x}\|^2, 
    \label{eq:cost}
\end{equation}
where $\eta\in\mathbb{R}$ and $\eta>0$, 
can be regarded as the regularized least square objective function for MMSE signal estimation 
because the unique stationary point of $f(\bm{x})$ coincides with 
the MMSE estimate \eqref{eq:mmse} when $\eta=\sigma^2$ \cite{Li}.
The scalar value $\eta$ in \eqref{eq:cost} behaves as a regularization parameter.
The gradient vector of $f(\bm{x})$ is given by 
\begin{equation}
    \nabla f(\bm{x})=(\bm{H}^\mathrm{H}\bm{H}+\eta\bm{I})\bm{x}-\bm{H}^\mathrm{H}\bm{y}.
\end{equation}

In this paper, 
we consider gradient flow \cite{Helmke,Strogatz} in terms of the objective function \eqref{eq:cost} 
for realizing continuous-time MMSE estimation configurable with analog optical devices.
{{Gradient flow illustrated in Fig.~\ref{fig:gradientflow} is a continuous-time counterpart of the discrete-time gradient descent method 
and is defined as $d\bm{x}(t)/dt=-\nabla f(\bm{x}(t))$ for a smooth function $f$.}}
Let $\bm{x}(t)\in\mathbb{C}^n$ be a continuous-time estimate of the transmitted signal $\bm{s}$ at time $t\geq 0$. 
By considering the gradient flow of \eqref{eq:cost}, we obtain the estimate $\bm{x}(t)$ that evolves 
according to the ODE 
\begin{equation}
    \frac{d\bm{x}(t)}{dt} =-\nabla f(\bm{x}(t)) 
    = -(\bm{H}^\mathrm{H}\bm{H}+\eta\bm{I})\bm{x}(t)+\bm{H}^\mathrm{H}\bm{y}.
    \label{eq:ode}
\end{equation}
We further assume the initial condition 
$\bm{x}(0) = \bm{H}^\mathrm{H}\bm{y}$, 
which corresponds to a matched filter detector \cite{Hama}.
The equilibrium point of the ODE \eqref{eq:ode} can be obtained 
as a solution to the equation $d\bm{x}(t)/dt=0$ 
and is given by $\bm{x}^\ast=(\bm{H}^\mathrm{H}\bm{H}+\eta\bm{I})^{-1}\bm{H}^\mathrm{H}\bm{y}$.
The point is unique 
due to the strict convexity of the potential function \eqref{eq:cost}, 
and coincides with the MMSE estimate when $\eta=\sigma^2$. 
We refer to the proposed signal detection method based on the ODE \eqref{eq:ode} as 
\emph{Ordinary Differential Equation-based MMSE (ODE-MMSE) method}.

\subsection{Implementation}
Implementation of the system \eqref{eq:ode} can be realized by using optical devices 
for computing products of matrix and vector.
The ODE-MMSE estimation system using the optical computing device is conceptually shown in Fig.~\ref{fig:analog}.
Electronic signals are converted to optical signals for a system combining an electronic computer with an optical computing device,
or, for an all-optical computer, all signals are optical \cite{Wu}.
The system is composed of analog adder, integrator, and matrix-vector multiplication device.
The most burdensome calculation is the matrix-vector multiplication $-(\bm{H}^\mathrm{H}\bm{H}+\eta\bm{I})\bm{x}(t)$ 
and can be delivered by using optical devices such as an array of MZIs.
The estimate $\bm{x}(t)$ is extracted after a sufficient amount of time has elapsed 
and then converted to detected symbol $\hat{\bm{s}}$ through symbol detection.

Analog optical devices for energy-efficient matrix multiplication are being developed \cite{Wu,mzi,Zhang,Lin} 
and such system performance is being experimentally demonstrated \cite{Prabhu2}, as discussed in Sect.~\ref{sec:intro}.
Recent works \cite{HZhang,Prabhu2} introduced specific circuit design using MZIs for matrix manipulation similar to the ODE-MMSE system.

\section{MSE Analysis}
\label{sec:analysis}
In this section, we derive an analytical formula for MSE 
obtained by the ODE-MMSE method.

\subsection{Properties}
In the following, the channel matrix $\bm{H}$ 
is assumed not to be a zero matrix, 
the mean of the transmitted signal is zero, 
and the second moment of the transmitted signal is assumed to be $\bm{I}$.

A closed-form representation of the estimate $\bm{x}(t)$ can be obtained 
using the solution for a first-order linear ODE with constant coefficients discussed in Sect.~\ref{sec:solution}.
The representation described below includes matrix inverse 
so that it requires the same computational burden as the original MMSE estimation \eqref{eq:mmse}. 
{{As noted in the introduction, 
this computational burden is severe 
due to the requirements of next generation wireless communication systems 
and hardware limitations.
Therefore, the implementation based on matrix-vector multiplication using optical devices according to \eqref{eq:ode} 
is necessary to achieve highly energy-efficient processing.}}
However, the closed-form representation provides analytical insights into the ODE-MMSE method, which will be discussed in the next subsection.
\begin{prop}
    The estimate of the ODE-MMSE method at time $t\geq0$ that follows ODE \eqref{eq:ode} is 
    given by 
    \begin{align}
        \bm{x}(t) = \bm{Q}(t)(\bm{Hs}+\bm{w}), 
        \label{eq:solution}
    \end{align}
    where 
    \begin{align}
        \bm{Q}(t) &:= \exp{(-(\bm{H}^\mathrm{H}\bm{H}+\eta\bm{I})t)}
        \left(\bm{I}-(\bm{H}^\mathrm{H}\bm{H}+\eta\bm{I})^{-1}\right)\bm{H}^\mathrm{H} \nonumber \\
        & \ + (\bm{H}^\mathrm{H}\bm{H}+\eta\bm{I})^{-1}\bm{H}^\mathrm{H}.
        \label{eq:Qt}
    \end{align}
\end{prop}
\textit{Proof}: 
The closed-form estimate $\bm{x}(t)$ can be obtained by deriving 
the analytical solution of the residual error vector between $\bm{x}(t)$ and the equilibrium point $\bm{x}^\ast$.
The residual error vector is defined as 
$\bm{e}(t):=\bm{x}(t)-\bm{x}^\ast$, 
and then the ODE \eqref{eq:ode} can be replaced with 
\begin{align}
    \frac{d\bm{e}(t)}{dt}  
    = -(\bm{H}^\mathrm{H}\bm{H}+\eta\bm{I})\bm{e}(t).
    \label{eq:et}
\end{align}
This is a typical first-order linear ODE with constant coefficients and 
can be solved using a matrix exponential (see Sect.~\ref{sec:solution}).
The solution is given by 
\begin{align}
    \bm{e}(t) &= \exp\left(-(\bm{H}^\mathrm{H}\bm{H}+\eta\bm{I})t\right)\bm{e}(0) \nonumber \\
    &= \exp\left(-(\bm{H}^\mathrm{H}\bm{H}+\eta\bm{I})t\right)\left(\bm{I}-(\bm{H}^\mathrm{H}\bm{H}+\eta\bm{I})^{-1}\right)\bm{H}^\mathrm{H} \bm{y}.
    \label{eq:eteq}
\end{align}
Therefore, the solution to \eqref{eq:ode} can be obtained 
by substituting \eqref{eq:eteq} and \eqref{eq:y} into $\bm{x}(t) = \bm{e}(t) + \bm{x}^\ast$, 
and by summarizing the terms in the equation.
\qed


The stability of the system \eqref{eq:et} can be evaluated 
via the eigenvalues of the matrix $\bm{A}:=\bm{H}^\mathrm{H}\bm{H}+\eta\bm{I}$.
\begin{prop}
    \label{prop:stability}
    The system \eqref{eq:et} is asymptotically stable.
\end{prop}
\textit{Proof}: 
From \eqref{eq:et}, the stability of the system depends on the Hermitian matrix $-\bm{A}=-(\bm{H}^\mathrm{H}\bm{H}+\eta\bm{I})$.
The Hermitian matrix $\bm{H}^\mathrm{H}\bm{H}$ is positive semidefinite 
and the matrix $\eta\bm{I}$ is positive definite.
The Hermitian matrix $-\bm{A}$ becomes negative definite 
so that it only has real and negative eigenvalues.
Thus, the system \eqref{eq:et} is proven to be asymptotically stable (Sect.~\ref{sec:solution}).
\qed

From Proposition~\ref{prop:stability}, the ODE-MMSE method has the following property.
\begin{prop}
    The ODE-MMSE method minimizes the objective function \eqref{eq:cost}.
\end{prop}
\textit{Proof}: 
The equilibrium point $\bm{x}^\ast$ is a unique point for minimizing the objective function \eqref{eq:cost} 
where the derivative equals zero.
From Proposition~\ref{prop:stability}, 
the estimate of the ODE-MMSE method is guaranteed to converge to the equilibrium point, i.e., the minimum value.
Therefore, the estimate of the ODE-MMSE method converges to a unique point to minimize the objective function.
\qed


\subsection{MSE Analysis}
\label{sec:derivation}
The MSE between the estimate $\bm{x}(t)$ and transmitted signal $\bm{s}$, 
\begin{equation}
    \mathrm{MSE}(t) := \mathbb{E}[\|\bm{x}(t)-\bm{s}\|^2], 
    \label{eq:originalmse}
\end{equation} 
is the principal performance indicator for MIMO signal detection methods \cite{Joham} 
but the analytical formula cannot always be derived \cite{Hayakawa}.
However, 
the proposed method has the advantage that 
the analytical formula for MSE can be described in a closed form 
without any constraints on system parameters, 
as shown in the following Theorem~\ref{theo:analyticalmse}.

In this section, we derive an analytical formula and asymptotic value of MSE using the eigenvalue decomposition of 
the Gram matrix $\bm{H}^\mathrm{H}\bm{H}$.
Suppose that the Gram matrix is decomposed as 
\begin{equation}
    \bm{H}^\mathrm{H}\bm{H} = \bm{U}\mathrm{diag}[\lambda_1,\ldots,\lambda_n]\bm{U}^\mathrm{H},
\end{equation}
where $\bm{U}\in\mathbb{C}^{m\times m}$ is a unitary matrix composed of the eigenvectors 
and $\lambda_1,\ldots,\lambda_n$ are nonnegative eigenvalues.
For convenience in subsequent analyses, we assume $\lambda_1\geq\ldots\geq\lambda_n\geq0$.
Condition number $\kappa$ of the Gram matrix is defined as $\kappa:=\lambda_1/\lambda_n$.
By using the decomposition, the following theorem holds:
\begin{theorem}
    \label{theo:analyticalmse}
    The MSE for the ODE-MMSE method is given by 
    \begin{align}
        \mathrm{MSE}(t) &= \sum_{i=1}^n \frac{\lambda_i(\lambda_i+\eta-1)^2(\lambda_i+\sigma^2)e^{-2(\lambda_i+\eta)t}}{(\lambda_i+\eta)^2} \nonumber \\
        &\ - \sum_{i=1}^n \frac{2\lambda_i(\lambda_i+\eta-1)(\eta-\sigma^2)e^{-(\lambda_i+\eta)t}}{(\lambda_i+\eta)^2} \nonumber \\
        &\ + \sum_{i=1}^n \frac{\eta^2+\sigma^2\lambda_i}{(\lambda_i+\eta)^2}.
        \label{eq:mse}
    \end{align}
\end{theorem}
\textit{Proof}: 
Substituting \eqref{eq:solution} into 
the right-hand side of \eqref{eq:originalmse} yields 
\begin{align}
	\mathrm{MSE}(t)
	&= \mathbb{E}\left[\|\left(\bm{Q}(t)\bm{H}-\bm{I}\right)\bm{s} 
    +\bm{Q}(t)\bm{w}\|^2 \right] \nonumber \\
	&= \mathrm{Tr}\left[(\bm{Q}(t)\bm{H}-\bm{I})^\mathrm{H}(\bm{Q}(t)\bm{H}-\bm{I})\right] \nonumber \\
	&\ + \sigma^2\mathrm{Tr}\left[\bm{Q}(t)^\mathrm{H}\bm{Q}(t)\right].
    \label{eq:tmpmse}
\end{align}
The matrix exponential $e^{-(\bm{H}^\mathrm{H}\bm{H}+\eta\bm{I})t}$ in $\bm{Q}(t)$
can be diagonalized using the eigenvalues of the Gram matrix as 
\begin{equation}
	e^{-(\bm{H}^\mathrm{H}\bm{H}+\eta\bm{I})t} = \bm{U}\mathrm{diag}[e^{-(\lambda_1+\eta)t}, \ldots, e^{-(\lambda_n+\eta)t}]\bm{U}^\mathrm{H}.
\end{equation}
Thus, the terms in \eqref{eq:tmpmse} can be diagonalized and calculated 
as 
\begin{align}
	&\mathrm{Tr}[\bm{Q}(t)^\mathrm{H}\bm{Q}(t)] = \sum_{i=1}^n \frac{\lambda_i\left(e^{-(\lambda_i+\eta)t}(\lambda_i+\eta-1)+1\right)^2}{(\lambda_i+\eta)^2} 
    \label{eq:tmp1}
\end{align}
and 
\begin{align}
	&\mathrm{Tr}\left[\left(\bm{Q}(t)\bm{H}-\bm{I}\right)^\mathrm{H}\left(\bm{Q}(t)\bm{H}-\bm{I}\right)\right] \nonumber \\
	&= \sum_{i=1}^n \frac{\left(\lambda_i(\lambda_i+\eta-1)e^{-(\lambda_i+\eta)t}-\eta\right)^2}{(\lambda_i+\eta)^2},
    \label{eq:tmp2}
\end{align}
respectively.
Detailed calculations are shown in Appendix~\ref{sec:appa}.
The MSE formula \eqref{eq:mse} is obtained by summarizing the terms of the matrix exponential.
\qed

We mention the MSE value of MMSE estimation \eqref{eq:mmse}.
\begin{lemma}
    The MSE of MMSE estimation \eqref{eq:mmse}, 
    $\mathrm{MSE}_\mathrm{mmse}:=\mathbb{E}[\|\hat{\bm{x}}-\bm{s}\|^2]$, is given by 
    \begin{equation}
        \mathrm{MSE}_\mathrm{mmse}=\sum_{i=1}^n \frac{\sigma^2}{\lambda_i+\sigma^2}.
        \label{eq:mseofmmse}
    \end{equation}
\end{lemma}
\textit{Proof}: 
This can be derived by using the MMSE estimate \eqref{eq:mmse} and 
the eigenvalue decomposition of the Gram matrix. 
The detailed derivation is provided in Appendix~\ref{sec:appb}.
\qed

Theorem~\ref{theo:analyticalmse} explicitly gives analytical MSE values 
of ODE-MMSE method at any time $t\geq0$.
By using this formula, we can describe the asymptotic MSE value, 
i.e., $\mathrm{MSE}(t)$ at the asymptotic limit of $t$. 
\begin{lemma}
    \label{lemma:asymptotic}
    Asymptotic MSE value for the ODE-MMSE method, 
    $\mathrm{MSE}_\infty:=\lim_{t\to\infty}\mathrm{MSE}(t)$, is given by 
    \begin{equation}
        \mathrm{MSE}_\infty = \sum_{i=1}^n \frac{\eta^2+\sigma^2\lambda_i}{(\lambda_i+\eta)^2}.
        \label{eq:asymptoticmse}
    \end{equation}
\end{lemma}
\textit{Proof}: 
When $t\to\infty$, 
the first and second terms of \eqref{eq:mse} vanish 
because $\lambda_i\geq0$ for $i=1,\ldots,n$ and $\eta>0$.
The remaining term is the asymptotic MSE value.
\qed

The inequality $\mathrm{MSE}_\mathrm{mmse}\leq\mathrm{MSE}_\infty$ holds and 
the equality holds if and only if $\eta=\sigma^2$ 
because the difference between \eqref{eq:asymptoticmse} and \eqref{eq:mseofmmse} 
\[
    \mathrm{MSE}_\infty-\mathrm{MSE}_\mathrm{mmse}
    = \sum_{i=1}^n\frac{\lambda_i(\eta-\sigma^2)^2}{(\lambda_i+\eta)^2(\lambda_i+\sigma^2)}
\]
is always nonnegative and equals $0$ if and only if $\eta=\sigma^2$.
This is consistent with the fact that 
the MMSE is the best linear estimator in terms of MSE \cite{mimobook}.

From Theorem~\ref{theo:analyticalmse} and Lemma~\ref{lemma:asymptotic}, 
we can find that the regularization parameter $\eta$ controls 
the convergence rate and asymptotic MSE value of the ODE-MMSE method.
The convergence rate depends significantly on the behavior of the exponential terms in \eqref{eq:mse}.
A larger value of $\eta$ accelerates the decrease in exponential terms 
but the asymptotic MSE value can be large because the value that minimizes the asymptotic MSE is achieved at $\eta=\sigma^2$.

\section{Time-dependent Regularization Parameter}
\label{sec:odetime}
This section introduces time-dependent control of the regularization parameter 
to improve the convergence property of the ODE-MMSE method.

According to the theoretical results in the previous section, 
the regularization parameter $\eta$ significantly affects 
the convergence properties of the ODE-MMSE method.
Theorem~\ref{theo:analyticalmse} and Lemma~\ref{lemma:asymptotic} indicate that 
a larger $\eta$ yields faster convergence of the ODE-MMSE method 
but yields a worse MSE value than the original MMSE estimation ($\mathrm{MSE}_{\mathrm{mmse}}$).
From these results, 
the adoption of time-dependent control of the regularization parameter $\eta$ is expected 
to hold both properties of faster convergence and a better asymptotic MSE value.
In this section, we improve the ODE-MMSE method 
to be more flexible 
by employing a time-dependent regularization parameter $\eta(t)$. 

We consider an estimate of $\bm{s}$ 
that evolves according to the following ODE: 
\begin{equation}
    \frac{d\bm{x}(t)}{dt} 
    = -(\bm{H}^\mathrm{H}\bm{H}+\eta(t)\bm{I})\bm{x}(t)+\bm{H}^\mathrm{H}\bm{y}.
    \label{eq:odetime}
\end{equation}
The expression $\eta(t)$ implies that 
the regularization parameter can vary depending on the time $t$.
The initial condition is the same as that in \eqref{eq:ode}, i.e., 
$\bm{x}(0) = \bm{H}^\mathrm{H}\bm{y}$.
We name the proposed method based on ODE \eqref{eq:odetime} 
\emph{ODE-MMSE with time-dependent regularization parameter (tODE-MMSE) method}.

The ODE \eqref{eq:odetime} can be analytically solved using the variation of parameters method \cite{numericalsolutionbook} 
because the matrix $\bm{A}(t):=\bm{H}^\mathrm{H}\bm{H}+\eta(t)\bm{I}$ is commutative.
\begin{prop}
    Estimate of the tODE-MMSE method at time $t\geq0$ that follows the ODE \eqref{eq:odetime} 
    is given by 
    \begin{align}
        \bm{x}(t) &= \exp{\left(-\bm{H}^\mathrm{H}\bm{H}t-\xi(t) \bm{I}\right)} \nonumber \\
        &\ \cdot\left(\bm{I}+ \int_0^t e^{\bm{H}^\mathrm{H}\bm{H}u+\xi(u) \bm{I}}du\right) \bm{H}^\mathrm{H}\bm{y},  
        \label{eq:solutiontime}
    \end{align}
    where $\xi(T):=\int_0^T \eta(s)ds$.
\end{prop}

Even in this case, the MSE formula for \eqref{eq:solutiontime} can be derived 
in the same manner as in Sect.~\ref{sec:derivation}.
\begin{theorem}
    The MSE for the tODE-MMSE method is given by 
    \begin{align}
        \mathrm{MSE}(t) &= \sum_{i=1}^n \lambda_i(\lambda_i+\sigma^2)\!\left(1\!+\!\int_0^t e^{\lambda_iu+\xi(u)}du\right)^2 \!e^{-2(\lambda_i t+\xi(t))} \nonumber \\
        &\ -2\sum_{i=1}^n\lambda_i\left(1+\int_0^t e^{\lambda_iu+\xi(u)}du\right) e^{-(\lambda_i t+\xi(t))}+n.
        \label{eq:msetime}
    \end{align}
    \label{theo:analyticalmsetime}
\end{theorem}
\textit{Proof}: 
MSE can be derived using the same procedure as in Theorem~\ref{theo:analyticalmse} 
by employing the eigenvalue decomposition of the Gram matrix.
Note that $\xi(t)$ is a scalar 
and that an integral in terms of a matrix is applied elementwise.
The detailed derivation is shown in Appendix~\ref{sec:appc}.
\qed

We obtain the result of Theorem~\ref{theo:analyticalmse} 
by setting $\eta(t)=\eta$.
The analytical formula \eqref{eq:msetime} has a complicated form, 
but, as with the ODE-MMSE method, 
the form of the time-dependent function $\eta(t)$ influences behavior of the estimation.

To obtain a better trade-off between convergence speed and asymptotic MSE value, 
it is considered better to choose $\eta(t)$ 
to become larger values in the transition phase before convergence 
and to set $\eta(t)=\sigma^2$ near convergence.
Furthermore, it is preferable that the integral $\xi(t)=\int_0^t \eta(s)ds$ is analytically tractable.
For example, we can use the following parametric models for the function $\eta(t)$: 
\begin{equation}
    \eta(t)=\frac{1}{\alpha t+\epsilon}+\sigma^2, 
    \label{eq:etainv}
\end{equation} 
\begin{equation}
    \eta(t) = \beta\exp{(-\gamma t)}+\sigma^2,
\end{equation}
where $\alpha,\beta,\gamma$ are positive parameters and $\epsilon$ is a small number fixed at $10^{-8}$ in this paper.
These functions converge to $\sigma^2$ at the limit $t\to\infty$.
Their integrals can be calculated as $\xi(t) = 1/\alpha\log\left((\alpha t+\epsilon)/\epsilon\right)+\sigma^2 t$ 
and $\xi(t)=\frac{\beta}{\gamma}(1-\exp{(-\gamma t)})+\sigma^2 t$, respectively.
The performance comparison with the ODE-MMSE method is reported in the later section, Sect.~\ref{sec:exptode}.

\section{Numerical Experiments}
\label{sec:sim}

\subsection{Overview of Experiments}
In this paper, all simulations were performed on Julia \cite{julia} and Python 
using a standard (non-optical) computer 
{{to emulate the continuous-time behavior of the proposed methods}}.
Sionna \cite{sionna}, a Python library, was used for generating channels and signals.
The behavior of the ODE was simulated using a numerical method with sufficient accuracy.
The detailed settings of the numerical method are discussed in Appendix~\ref{sec:appd}.

The channel matrix $\bm{H}$ was generated in most subsequent experiments 
so that each element followed an independent and identically distributed $\mathcal{CN}(0,1)$, 
unless otherwise noted.

\subsection{ODE-MMSE}
Numerical examples are presented to confirm the validity of the MSE formula \eqref{eq:mse} 
and evaluate the impact of the parameter $\eta$ on the convergence rate and asymptotic MSE value \eqref{eq:asymptoticmse}.

\begin{figure}[tb]
    \centerline{\includegraphics[width=\columnwidth]{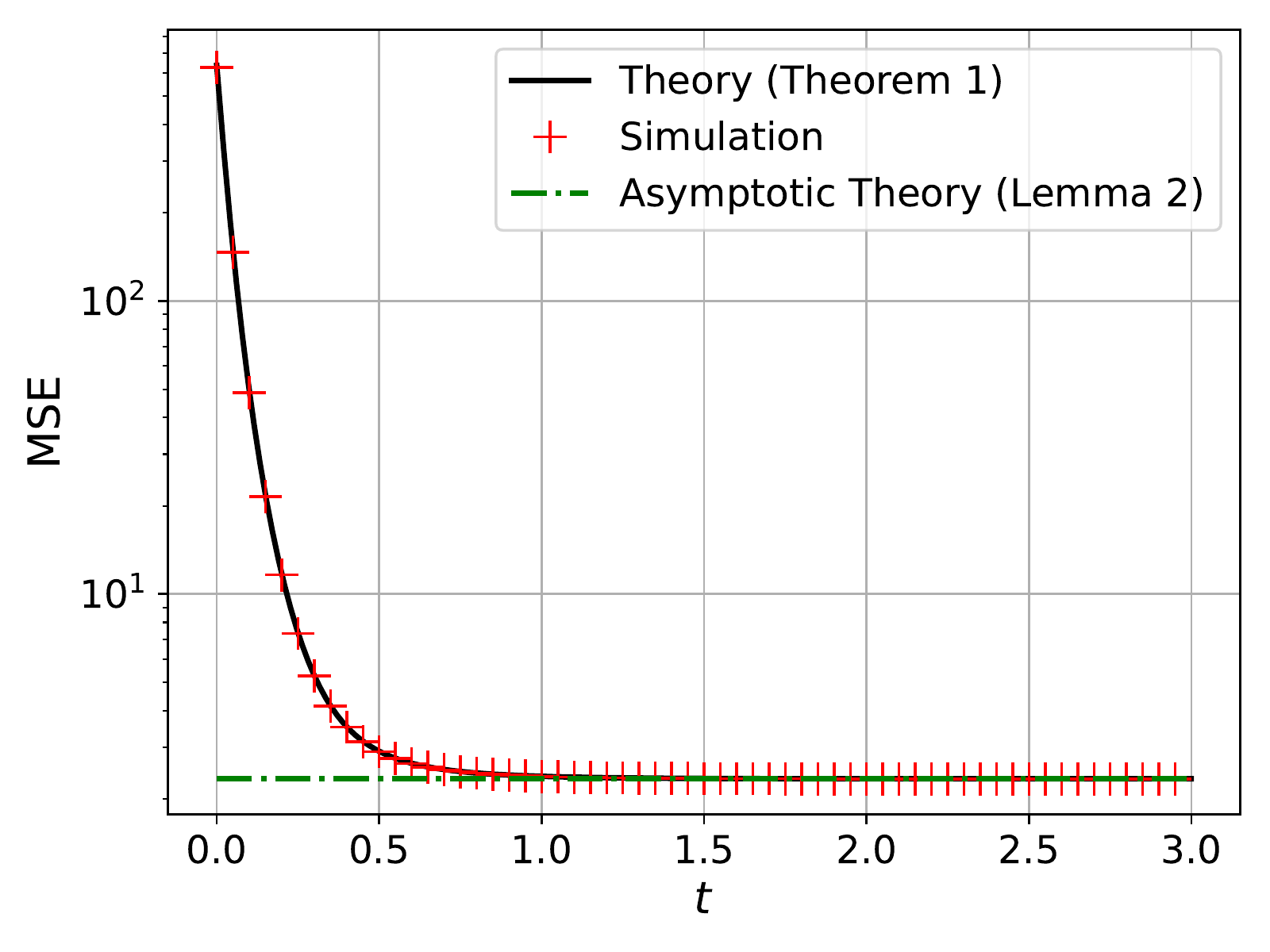}}
    \caption{MSE derived from analytical formula and calculated by simulation, $(n,m,\sigma^2,\eta,\kappa)=(8,8,1,0.5,164.17)$, QPSK.}
    \label{fig:Euler1}
\end{figure}%
\begin{figure}[tb]
    \centerline{\includegraphics[width=\columnwidth]{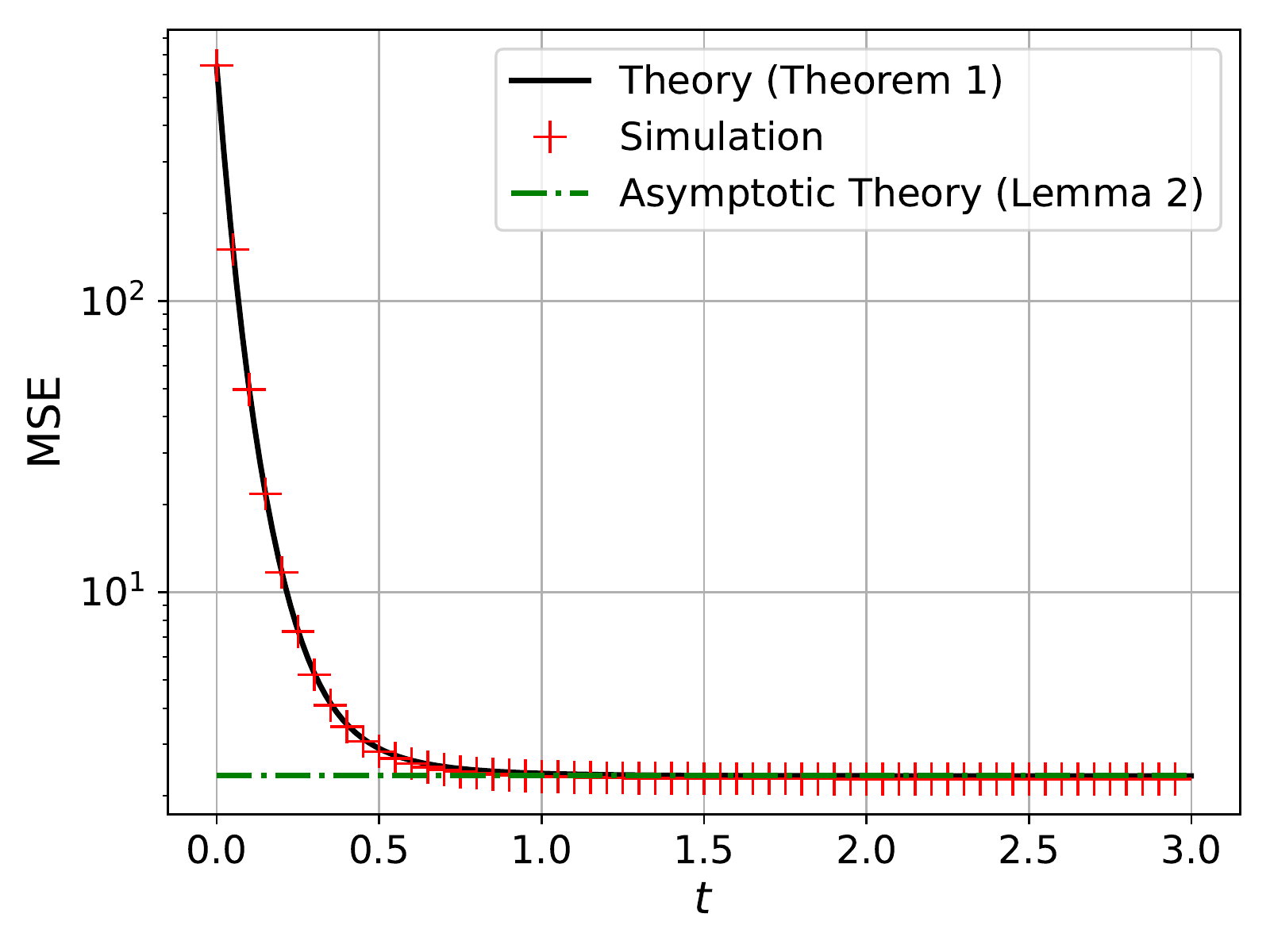}}
    \caption{MSE derived from analytical formula and calculated by simulation, $(n,m,\sigma^2,\eta,\kappa)=(8,8,1,0.5,164.17)$, 64QAM.}
    \label{fig:Euler2}
\end{figure}
\subsubsection{Confirmation of Analysis}
We verify the MSE obtained from Theorem~\ref{theo:analyticalmse} \eqref{eq:mse} for a single instance of the channel matrix $\bm{H}$. 
The condition number of the Gram matrix was $\kappa=164.17$.
Figures~\ref{fig:Euler1} and \ref{fig:Euler2} show the MSE values derived from the theorem \eqref{eq:mse}, 
those obtained from the simulation using the numerical method, 
and the asymptotic MSE value \eqref{eq:asymptoticmse} 
in the cases of QPSK and 64QAM transmitted signals, respectively.
The system parameters were set to $(n,m,\sigma^2,\eta)=(8,8,1,0.5)$.
The curve of the analytical formula is comparable to that of the simulation 
with sufficient accuracy.
We can see that the MSE converges to the asymptotic MSE value.
These results are consistent with the MSE analysis in Sect.~\ref{sec:derivation} 
where the MSE of the ODE-MMSE method can be described by the analytical formula \eqref{eq:mse}, 
and the MSE value asymptotically converges to the value given in \eqref{eq:asymptoticmse}.

\subsubsection{Influence of Regularization Parameter}
\begin{figure}[tbp]
    \centerline{\includegraphics[width=\columnwidth]{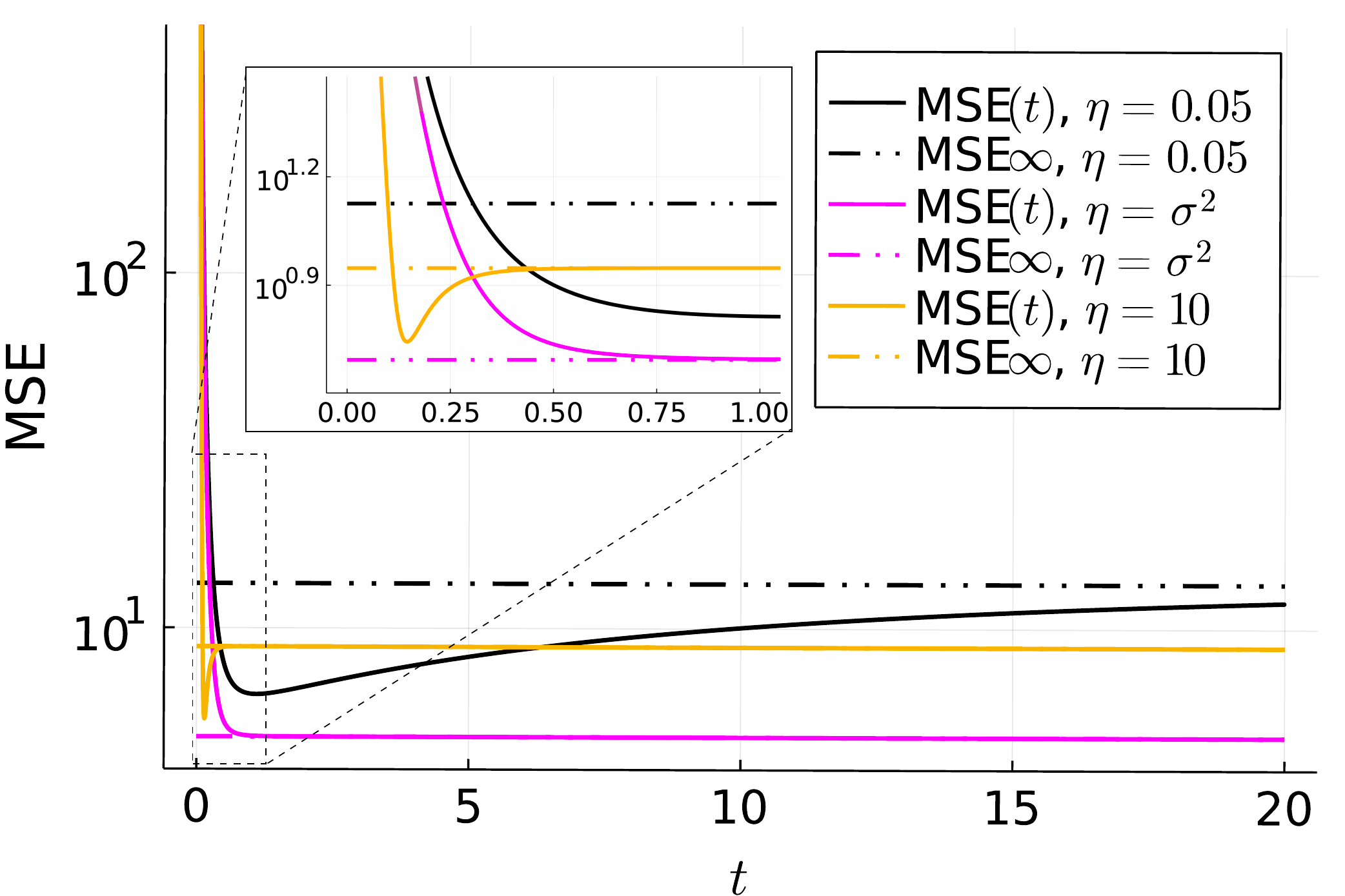}}
    \caption{Comparison of MSE with different choices of the regularization parameter $\eta$, $(n,m,\sigma^2,\kappa)=(32,32,1,3727.67)$.}
    \label{fig:compareeta}
\end{figure}
We also evaluated the influence of regularization parameter $\eta$ on the convergence behavior of the ODE-MMSE method.
Figure~\ref{fig:compareeta} shows the MSE values obtained from \eqref{eq:mse} for different values of $\eta$: $\eta=0.05,\sigma^2,10$.
The system parameters were set to $(n,m,\sigma^2)=(32,32,1)$.
The condition number of the Gram matrix was $\kappa=3727.67$.
Concerning the convergence rate, 
the MSE with $\eta=10$ decreases rapidly, 
and that with $\eta=0.05$ is the slowest among the choices.
This result is consistent with the interpretation of \eqref{eq:mse} 
where a larger $\eta$ accelerates the decay of the exponential terms.
On the other hand, for the asymptotic MSE values, 
the value is the lowest when $\eta=\sigma^2$ and 
the choice $\eta=0.005$ leads to the highest value although 
the MSE is lower for $0.5<t<1$ than that for $\eta=10$.
Therefore, 
the convergence behavior largely depends on the choice of regularization parameter $\eta$ 
and the superiority and inferiority of the MSE values can be switched depending on the time of interest.

\subsection{tODE-MMSE}
\label{sec:exptode}
We show numerical examples to confirm validity of the MSE formula \eqref{eq:msetime} 
and to compare the convergence performance of the tODE-MMSE method 
with that of the ODE-MMSE method.

\subsubsection{Confirmation of Analysis}
\begin{figure}[tb]
    \centerline{\includegraphics[width=\columnwidth]{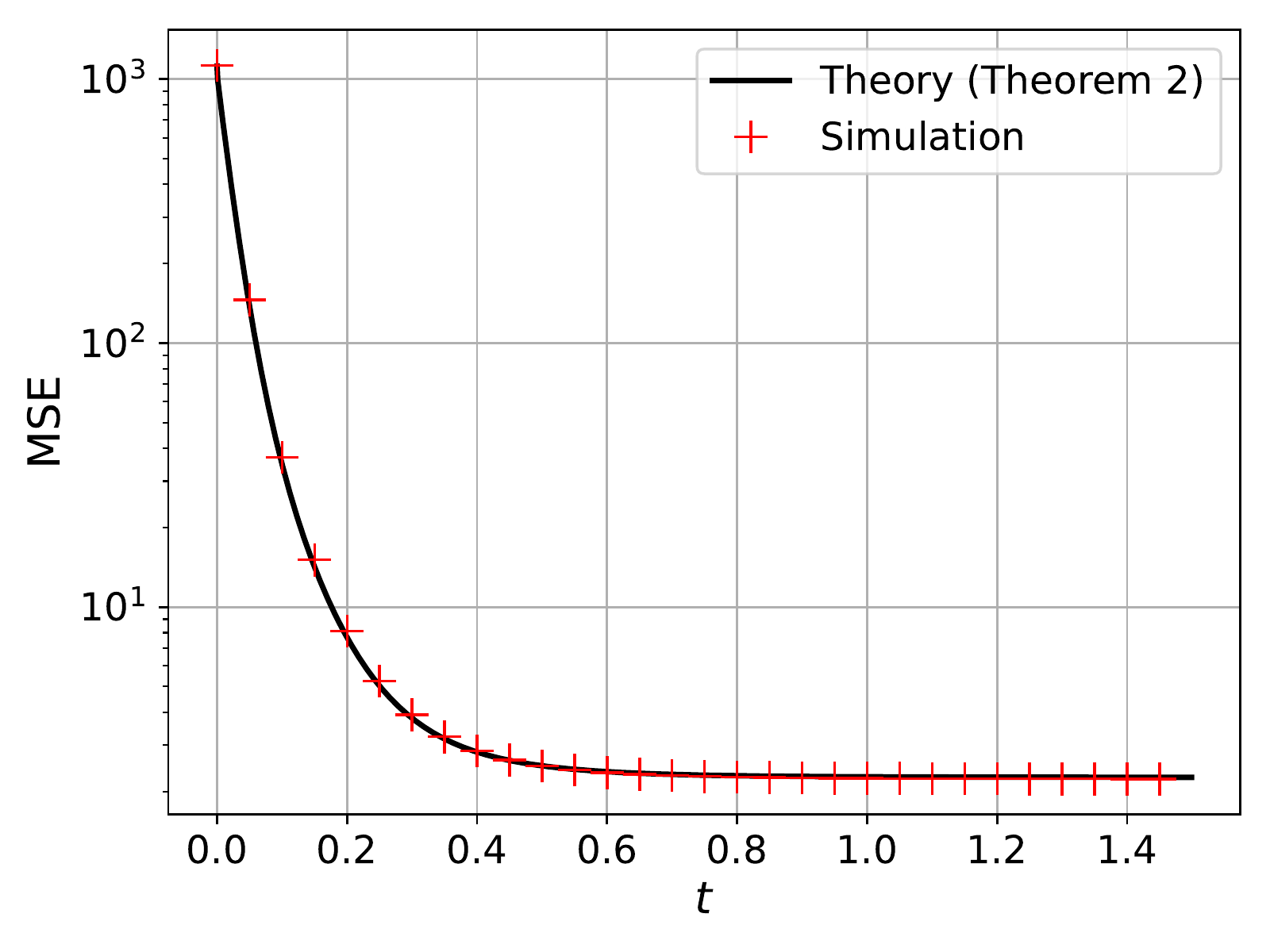}}
    \caption{MSE derived from analytical formula and calculated by simulation, $(n,m,\sigma^2,\alpha,\kappa)=(8,8,1,500,229.74)$, QPSK.}
    \label{fig:Euleretat1}
\end{figure}%
\begin{figure}[tb]
    \centerline{\includegraphics[width=\columnwidth]{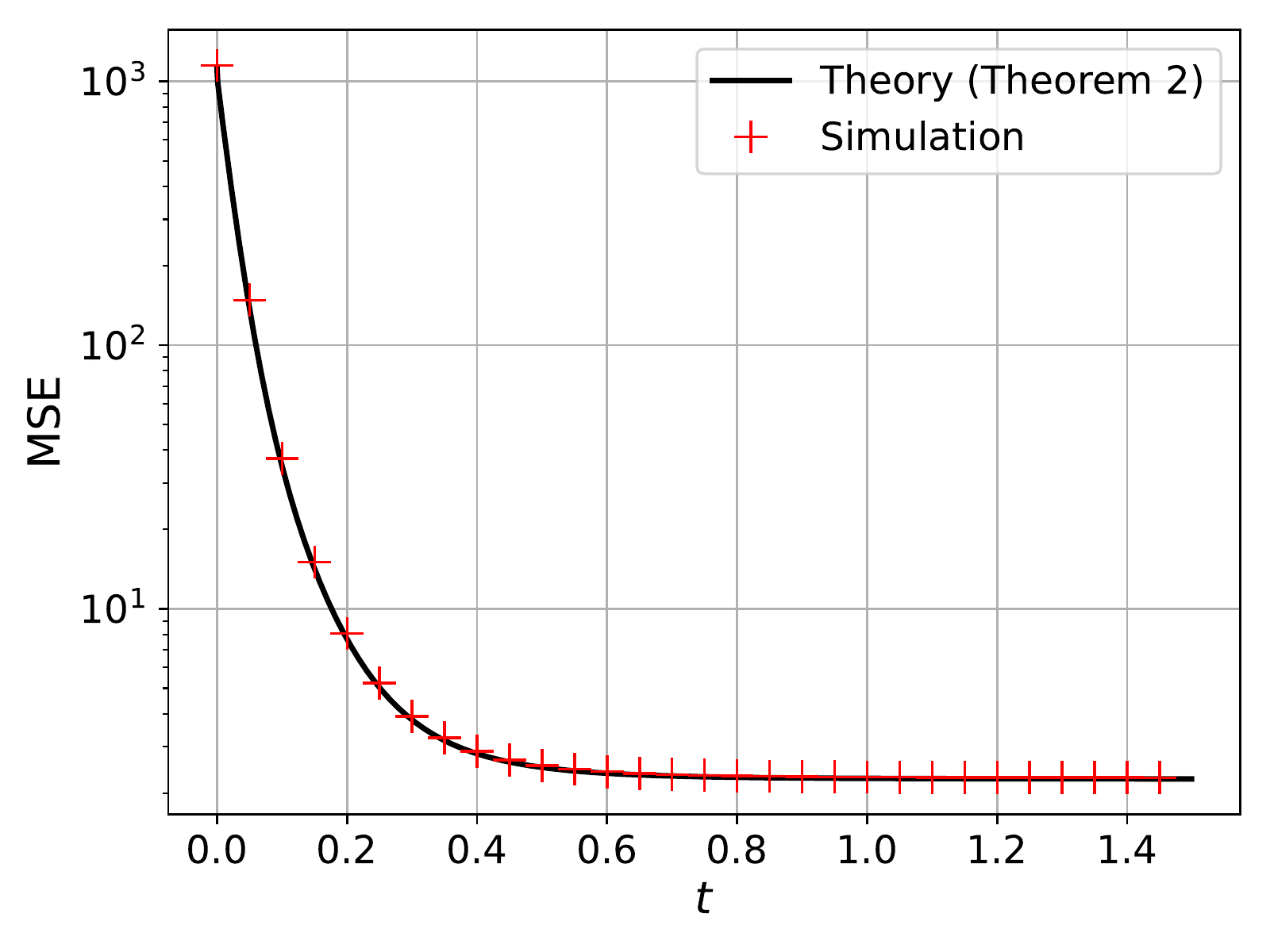}}
    \caption{MSE derived from analytical formula and calculated by simulation, $(n,m,\sigma^2,\alpha,\kappa)=(8,8,1,500,229.74)$, 64QAM.}
    \label{fig:Euleretat2}
\end{figure}
We verify the MSE obtained from Theorem~\ref{theo:analyticalmsetime} \eqref{eq:msetime} for a single instance of the channel matrix $\bm{H}$.
Figures~\ref{fig:Euleretat1} and \ref{fig:Euleretat2} show the MSE values at time $t$ of the methods with QPSK and 64QAM transmitted signals, respectively.
We used the tractable regularization function \eqref{eq:etainv} as the function $\eta(t)$ with $\alpha=500$.
The system parameters were set to $(n,m,\sigma^2)=(8,8,1)$.
The condition number of the Gram matrix was $\kappa=229.74$.
The curve of the MSE obtained using \eqref{eq:msetime} is comparable to that of the simulation with sufficient accuracy. 
This result is consistent with the MSE analysis in Sect.~\ref{sec:odetime}.

\subsubsection{Comparison of tODE-MMSE with ODE-MMSE}
We present an example that uses the MSE formula \eqref{eq:msetime} of the tODE-MMSE method 
for improving the convergence properties and 
compare the performance with that of the ODE-MMSE method.
We have found in Fig.~\ref{fig:compareeta} that the performance of the proposed method 
largely depends on the choice of regularization parameter. 
It is expected that 
we can improve the convergence property by the tODE-MMSE method with an appropriate choice of the function $\eta(t)$.
There are various possible indicators for evaluating the goodness of convergence performance.
In this paper, we employed a functional 
\[F(\xi(t)) := \int_0^T \mathrm{MSE}(t) dt\] 
as the indicator.
If a method has faster convergence and lower errors, 
the value of the functional decreases.
In the following, 
we optimize the parameter by minimizing the functional value.
Specifically, we employ a grid search to select the optimal parameter.

\begin{figure}[tbp]
    \centerline{\includegraphics[width=\columnwidth]{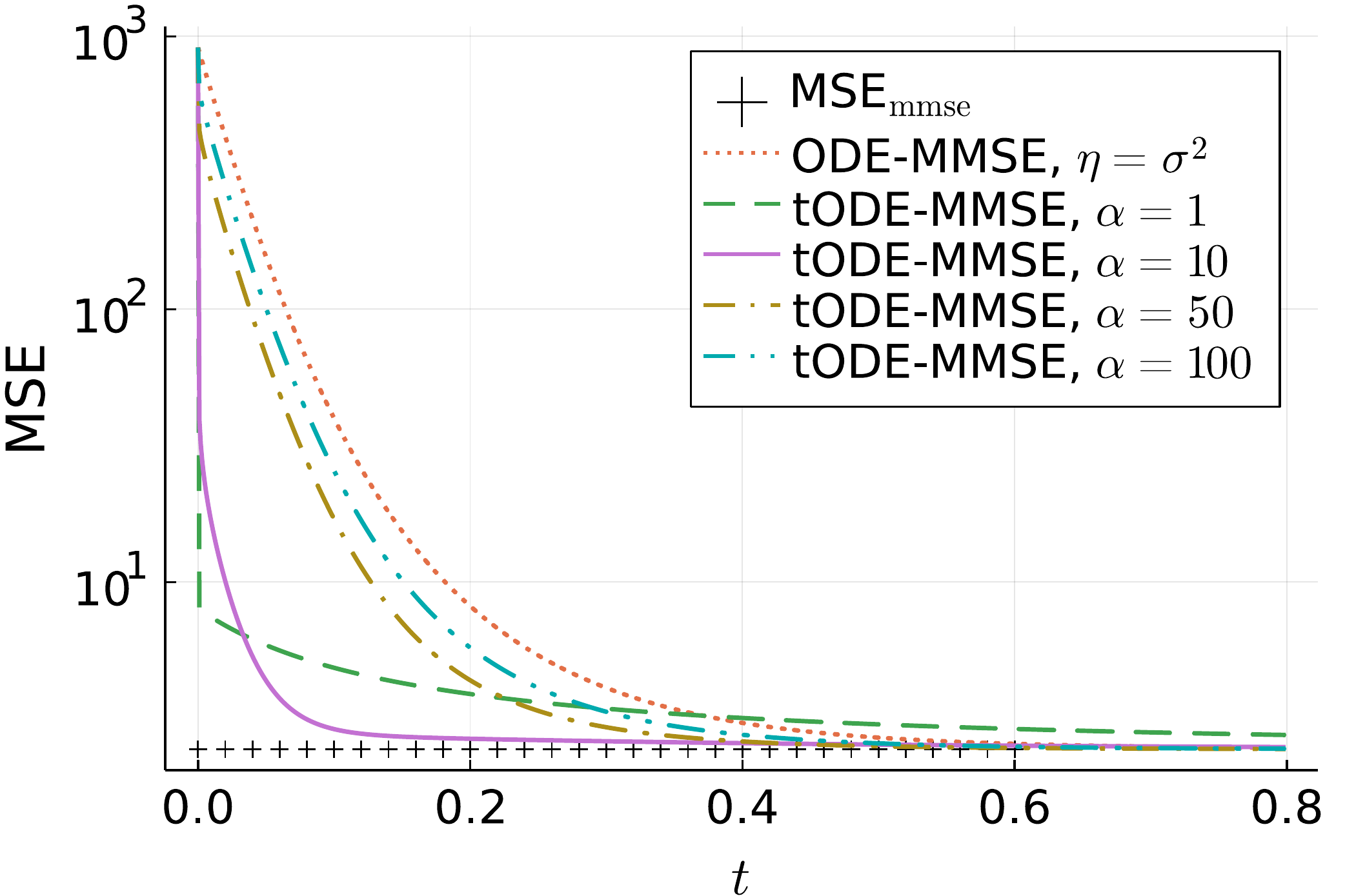}}
    \caption{The MSE curves with different values of $\alpha$, $(n,m,\sigma^2,\kappa)=(8,8,1,305.45)$.}
    \label{fig:gridsearch}
\end{figure}

We set $\alpha=1,10,50$, and $100$ as the parameter candidates.
The system parameters were set to $(n,m,\sigma^2)=(8,8,1)$ and $T=0.8$.
The condition number of the Gram matrix was $\kappa=305.45$.
Table~\ref{tab:tab1} summarizes the evaluated values of $F(\xi(t))=F(\alpha)$.
From the table, the value is found the lowest when $\alpha=10$. 
Fig.~\ref{fig:gridsearch} shows the MSE of MMSE estimate $\mathrm{MSE}_\mathrm{mmse}$, 
the MSE values of the ODE-MMSE method with $\eta=\sigma^2$, 
and those of the tODE-MMSE method for different values of $\alpha$.
From Fig.~\ref{fig:gridsearch}, 
most of the MSE curves of the tODE-MMSE method except for the case $\alpha=1$ 
converge to the value of $\mathrm{MSE}_\mathrm{mmse}$ 
faster than the ODE-MMSE method.
Furthermore, the method with $\alpha=10$, 
which has the lowest functional value in Table~\ref{tab:tab1}, 
exhibited the fastest convergence.
This indicates that an improved estimation method can be determined through a grid search using the functional value.
\begin{table}[tbp]
    \caption{Values of functional $F(\alpha)$.}
    \begin{center}
    \begin{tabular}{|c||c|c|c|c|c|}
        \hline
        $\alpha$ & $1$ & $10$ & $50$ & $100$ \\
        \hline
        $F(\alpha)$ & $2.8963$ & $2.5593$ & $13.8035$ & $19.5093$ \\
        \hline
    \end{tabular}
    \label{tab:tab1}
    \end{center}
\end{table}

\section{Continuous Time to Discrete Time}
\label{sec:discrete}
This section presents byproducts of considering the continuous-time system. 
By using the ODE perspective, 
it is possible to derive \emph{discrete-time} detection algorithms 
that run on electronic computers used today.
Moreover, analyses in the continuous-time estimation provide new insights into the discrete-time algorithms.

\subsection{Brief Review of Discrete-time Detection Algorithms}
\label{sec:discreteintro}
For massive MIMO systems in the next-generation communication systems, 
the computational burden of signal detection is an inescapable problem.
Maximum likelihood (ML) detection can achieve optimal performance 
but it requires exponentially increasing complexity with the number of antennas.
ZF and linear MMSE \cite{mimobook} are the simplest methods that require cubic complexity 
for matrix inverse calculation.
An iterative detection algorithm based on approximate message passing (AMP) \cite{Jeon} shows better performance than ZF and MMSE 
as long as the channel is well modeled by a particular probability distribution.
Orthogonal AMP (OAMP)-based detection \cite{oamp} relaxes this constraint 
and is suitable for a wider range of communication environments.
However, the computational complexity is higher than the above methods 
due to the inverse calculation per iteration.
Recently, detection algorithms based on deep learning have been proposed \cite{Khani,Pratik,He}.
He et al. proposed OAMP-Net2 \cite{He}, which is the combination of OAMP-based detection and deep learning 
and requires an additional cost for the learning process.

{{Performance of MMSE is far from that of the optimal ML 
and inferior to the state-of-the-art algorithms, 
especially in the cases of higher-order modulation and realistic channels \cite{Khani}.}}
However, reducing the computational complexity of MMSE still remains a topic being addressed 
because it remains the most basic detection method.
There are several approaches on the basis of iterative matrix-vector multiplication \cite{Gao,Dai,Minango,jsor}, 
which is preferred over inverse matrix calculation due to its suitability for parallel computing \cite{Prabhu}.
Naceur proposed an iterative MMSE estimation algorithm \cite{jsor} 
based on the Jacobi \cite{jacobi} and successive over-relaxation (SOR) methods \cite{sor}.
We call the method in this paper the Jacobi SOR algorithm.
The Jacobi SOR algorithm has been reported to achieve fast convergence in well-conditioned channels with $\mathcal{O}(n^2)$ complexity 
and have high parallelism.

\subsection{Discrete-time Algorithm from Discretization of ODE}
In this subsection, we derive discrete-time algorithms for MMSE estimation 
from the ODE perspective.
In the continuous-time ODE-MMSE method, the estimate of the transmitted signal 
evolves according to the ODE 
\begin{equation}
    \frac{d\bm{x}(t)}{dt} 
    = -(\bm{H}^\mathrm{H}\bm{H}+\eta\bm{I})\bm{x}(t)+\bm{H}^\mathrm{H}\bm{y}.
    \tag{\ref{eq:ode}}
\end{equation}
The behavior can be discretized and traced using numerical methods 
such as the well-known explicit Euler method and the 4th-stage (4th-order) Runge-Kutta method \cite{numericalsolutionbook}. 
In other words, discrete-time detection algorithms can be derived 
by discretizing this ODE.

For example, 
applying the explicit Euler method to the ODE yields the following update equation: 
\begin{equation}
    \bm{x}^{[k]}=\bm{x}^{[k-1]} -\delta(\bm{H}^\mathrm{H}\bm{H}+\eta\bm{I})\bm{x}^{[k-1]} + \delta\bm{H}^\mathrm{H}\bm{y}, 
    \label{eq:euler}
\end{equation}
where $\bm{x}^{[k]} \ (k=1,2,\ldots)$ is a discrete-time estimate of the transmitted signal at $k$th iteration 
and $\delta>0$ is the step-size parameter.
However, there is no novelty in applying the Euler method for the discretization 
because the update equation has the same formulation as that of the conventional estimation method based on the standard gradient descent method \cite{Berthe}.
There is also no advantage of using the 4th-stage Runge-Kutta method 
because the convergence performance corresponding to the required computational costs is comparable to the Euler method \cite{Riha}.

Adoption of more stable and computationally reasonable numerical methods \cite{numericalsolutionbook,Howen,Verwer,Riha,Eftekhari,Ushiyama} 
has a possibility of producing efficient detection algorithms.
The meaning of more stable methods is that we can employ a larger step-size parameter under guarantees of stability 
and it leads to a faster convergence of algorithms.
The Runge-Kutta Chebyshev descent (RKCD) method \cite{Eftekhari} is one of such numerical methods 
that has both stability and computational tractability with a computational cost comparable to that of the explicit Euler method. 
The RKCD method can achieve faster convergence than the standard gradient-based (Euler) method with comparable costs due to its high stability.
{{The application to MIMO detection has not been considered, to the best of our knowledge.}}

We can obtain a novel MMSE detection algorithm by applying the RKCD method to the ODE \eqref{eq:ode}.
The main update equation of the RKCD method is obtained by applying the Chebyshev polynomial \cite{Chebyshev} to 
the update equation of the $s$-stage Runge-Kutta method to introduce a flexible step size.
The update equation is given by 
\begin{equation}
    \bm{x}^{[k]} = -h\mu_j\nabla f(\bm{x}^{[k-1]}) + \nu_j \bm{x}^{[k-1]} - (\nu_j-1)\bm{x}^{[k-2]}, 
\end{equation}
where $j = \mod{(k-1,s)}+1$, 
\begin{equation}
    \mu_j := \frac{2\omega_1T_{j-1}(\omega_0)}{T_j(\omega_0)}, \ \nu_j := \frac{2\omega_0T_{j-1}(\omega_0)}{T_j(\omega_0)},
\end{equation}
and the parameters $s, \omega_0, \omega_1$, and $h$ are required to guarantee stability. 

The parameter $s$ can be regarded as the number of internal stages in the Runge-Kutta method, 
$\omega_0$ determines the stability region, $\omega_1$ is required to satisfy the consistency, 
and $h$ is a step-size parameter.
A reasonable choice of $\omega_0$ and $\omega_1$ \cite{Verwer} is known as 
\begin{equation}
    \omega_0 := 1+\frac{\epsilon}{s^2}, \ \omega_1 := \frac{T_s(\omega_0)}{T_s'(\omega_0)}, 
\end{equation}
where $\epsilon>0$ is a damping constant.
The parameters $s$ and $h$ can be chosen arbitrarily 
as long as the stability holds, 
and one choice guaranteeing the stability discussed in \cite{Eftekhari} is  
\begin{equation}
    s:=\left\lceil\sqrt{\frac{\left(-1+\frac{L}{\ell}\right)\epsilon}{2}}\right\rceil, \ 
    h := \frac{\omega_0-1}{\omega_1 (\ell+\eta)}, 
    \label{eq:sandh}
\end{equation}
where $\ell$ and $L$ are the lower and upper bounds of the eigenvalues of $\bm{H}^\mathrm{H}\bm{H}$, 
or simply the lowest and highest eigenvalues $\lambda_n$, $\lambda_1$, respectively.

Algorithm~\ref{alg:rkc} summarizes the detailed process of the RKCD method-based detection.
Computational complexity of this method is dominated by the matrix-vector product.
A performance comparison of the RKCD method-based MMSE estimation with recent detection methods (OAMP-based detection \cite{oamp} and OAMP-Net2 \cite{He}) 
is shown in Appendix~\ref{sec:appe}.
\begin{algorithm}[tb]
	\caption{RKCD method-based MMSE detection}
	\label{alg:rkc}
	\begin{algorithmic}[1]
		\REQUIRE Damping constant $\epsilon$, lower $\ell$ and upper bound $L$ for eigenvalues of $\bm{H}^\mathrm{H}\bm{H}$, initial value $\bm{x}_0$
		\ENSURE Estimated Symbol $\hat{\bm{s}}$
        \STATE Set parameters $s,h,\omega_0,\omega_1$
        \STATE $\bm{x}^{[0]}=\bm{x}_0$
        \FOR{$k=1,\ldots,J$}
            \IF{$\mod{(k,s)}==1$}
                \STATE $\bm{x}^{[k]} = \bm{x}^{[k-1]}-\frac{h\omega_1}{\omega_0}\left((\bm{H}^\mathrm{H}\bm{H}+\eta\bm{I})\bm{x}^{[k-1]}-\bm{H}^\mathrm{H}\bm{y}\right)$
            \ELSE
                \STATE $j = \mod{(k-1,s)}+1$
                \STATE $\mu_j = \frac{2\omega_1T_{j-1}(\omega_0)}{T_j(\omega_0)}$
                \STATE $\nu_j = \frac{2\omega_0T_{j-1}(\omega_0)}{T_j(\omega_0)}$
                \STATE $\bm{x}^{[k]} = -h\mu_j\left((\bm{H}^\mathrm{H}\bm{H}+\eta\bm{I})\bm{x}^{[k-1]}-\bm{H}^\mathrm{H}\bm{y}\right) + \nu_j \bm{x}^{[k-1]} - (\nu_j-1)\bm{x}^{[k-2]}$
            \ENDIF
        \ENDFOR
        \STATE Return $\hat{\bm{s}}$ by symbol detection using $\bm{x}^{[J]}$
	\end{algorithmic}
\end{algorithm}

\subsection{MSE Analysis for Discrete-time Algorithms}
There is considerable interest in understanding the relationship 
between performance and runtime of discrete algorithms.
For this purpose, it is desirable to clarify MSE performance in each iteration of the algorithms 
but the performance cannot always be traced analytically, as discussed in Sect.~\ref{sec:derivation}.
However, the use of ODE may facilitate this.

Solution of continuous-time methods represented by ODE can be numerically traced 
using numerical methods discretized with sufficient accuracy.
In this case, the MSE of the discrete-time numerical methods should agree with 
the result of the MSE analysis for the continuous-time method.
Therefore, the MSE analysis in continuous-time methods 
also allows us to analyze the performance of discrete-time algorithms obtained by discretizing the ODE.
In other words, the MSE performance of Algorithm~\ref{alg:rkc} obtained from ODE \eqref{eq:ode} 
can be described using the MSE formula \eqref{eq:mse}.
Note that the above discussion does not hold if one uses a numerical method with insufficient accuracy, i.e., with a large step-size parameter.


The estimate $\bm{x}^{[k]}$ at the $k$th iteration of Algorithm~\ref{alg:rkc} corresponds to 
the estimate at the following time $T_k$ ($k=1,2,\ldots$).
The time $T_k$ can be described recursively \cite{Howen} as 
\begin{equation}
    T_k \!= \!\begin{cases}
        \tilde{t}_k \!+\! \frac{h\omega_1}{\omega_0}, & \mathrm{if}\mod(k,s)=1, \\
        \tilde{t}_k \!+\! \nu_kT_{k-1}\!+\!(1\!-\!\nu_k)T_{k-2}\!+\!h\mu_k, & \mathrm{otherwise}, 
    \end{cases}
\end{equation}
where $\tilde{t}_0=0$ and $\tilde{t}_k = \tilde{t}_{k-1}+T_k$ if $\mod(k,s)=0$ otherwise, $\tilde{t}_k=\tilde{t}_{k-1}$.
This implies that the step-size parameters of the RKCD method depend on the iteration index.
The MSE value of the method, $\mathrm{MSE}_\mathrm{RKCD}[k]$, is given by 
\begin{equation}
    \mathrm{MSE}_{\mathrm{RKCD}}[k] \simeq \mathrm{MSE}(T_k).
\end{equation}



\subsection{Simulation Results}
\subsubsection{Comparison with Conventional MMSE Detection Method}
In this section, we evaluate the estimation performance of the discrete-time MMSE estimation algorithms 
obtained using numerical methods.
The performance is compared with that of the conventional Jacobi SOR algorithm \cite{jsor}, 
{{which does not require learning as well as the proposed algorithm}}.

\begin{figure}[tb]
    \begin{tabular}{cc}
      \begin{minipage}[b]{0.45\hsize}
        \centering
        \includegraphics[width=\columnwidth]{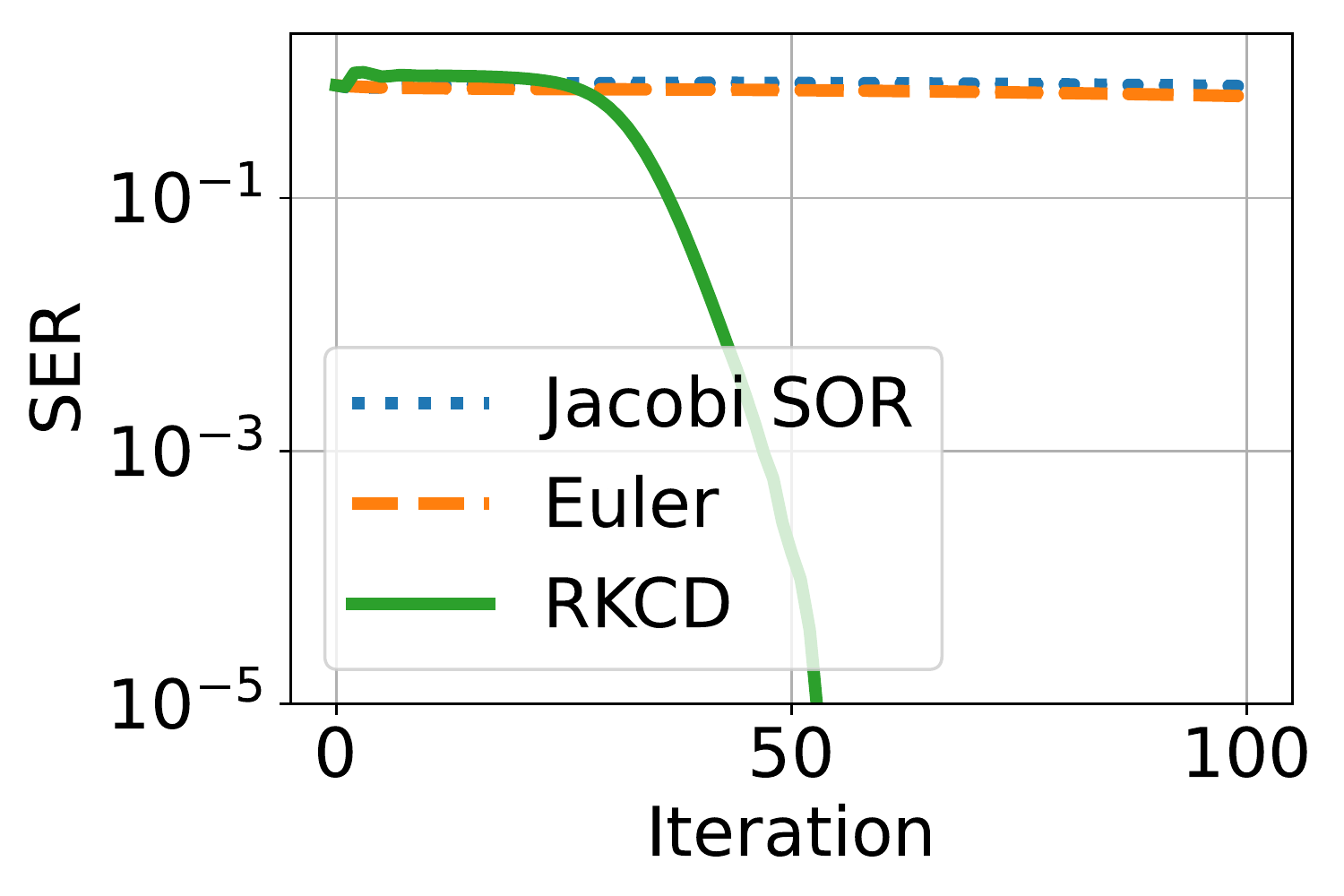}
      \end{minipage} & 
      \begin{minipage}[b]{0.45\hsize}
        \centering
        \includegraphics[width=\columnwidth]{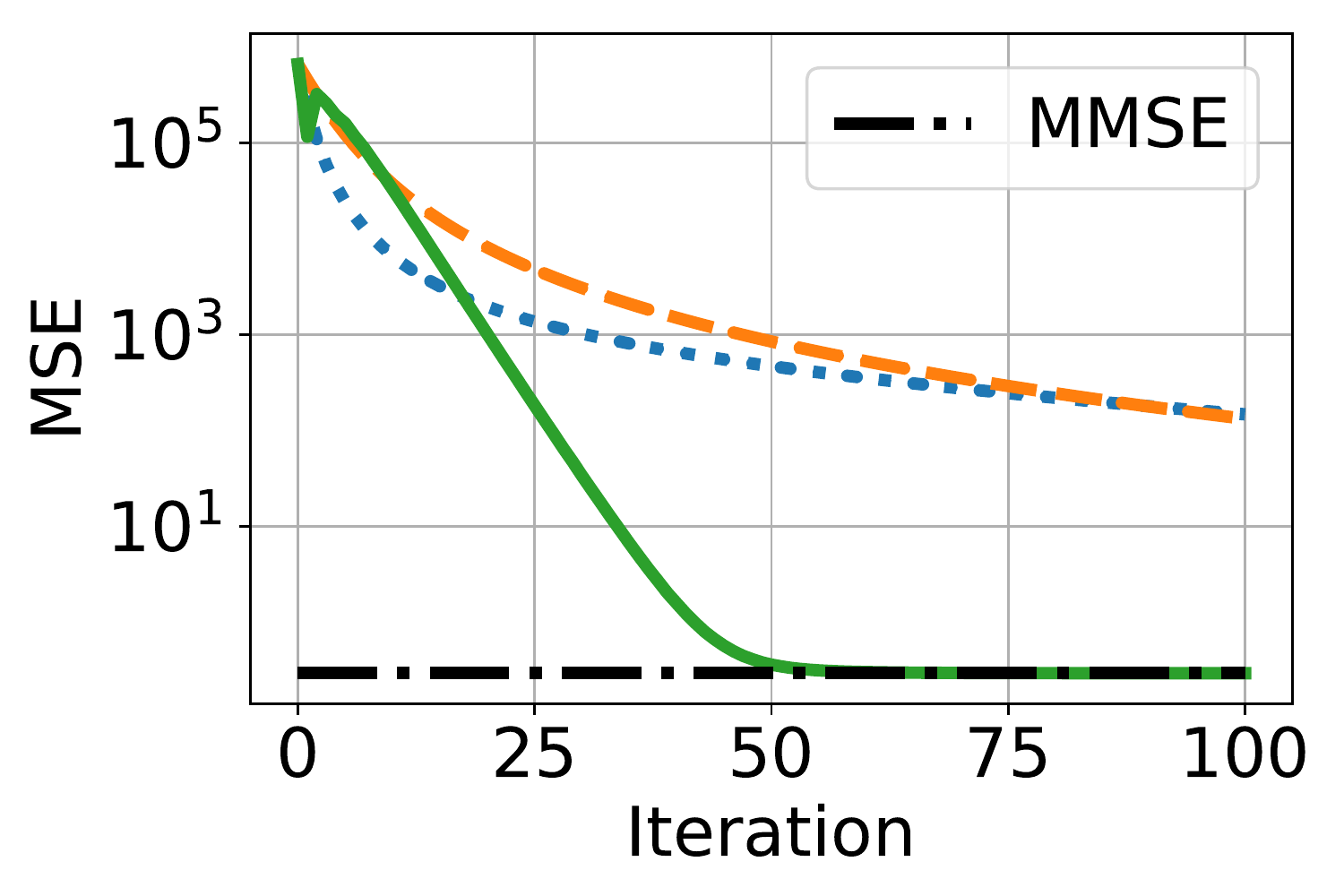}
      \end{minipage}
    \end{tabular}
    \caption{Comparison of discrete-time MMSE estimation methods, $(n,m,\sigma^2)=(60,80,0.1)$, 16QAM.}
    \label{fig:rkc}
\end{figure}
\begin{figure}[tb]
    \begin{tabular}{cc}
      \begin{minipage}[b]{0.45\hsize}
        \centering
        \includegraphics[width=\columnwidth]{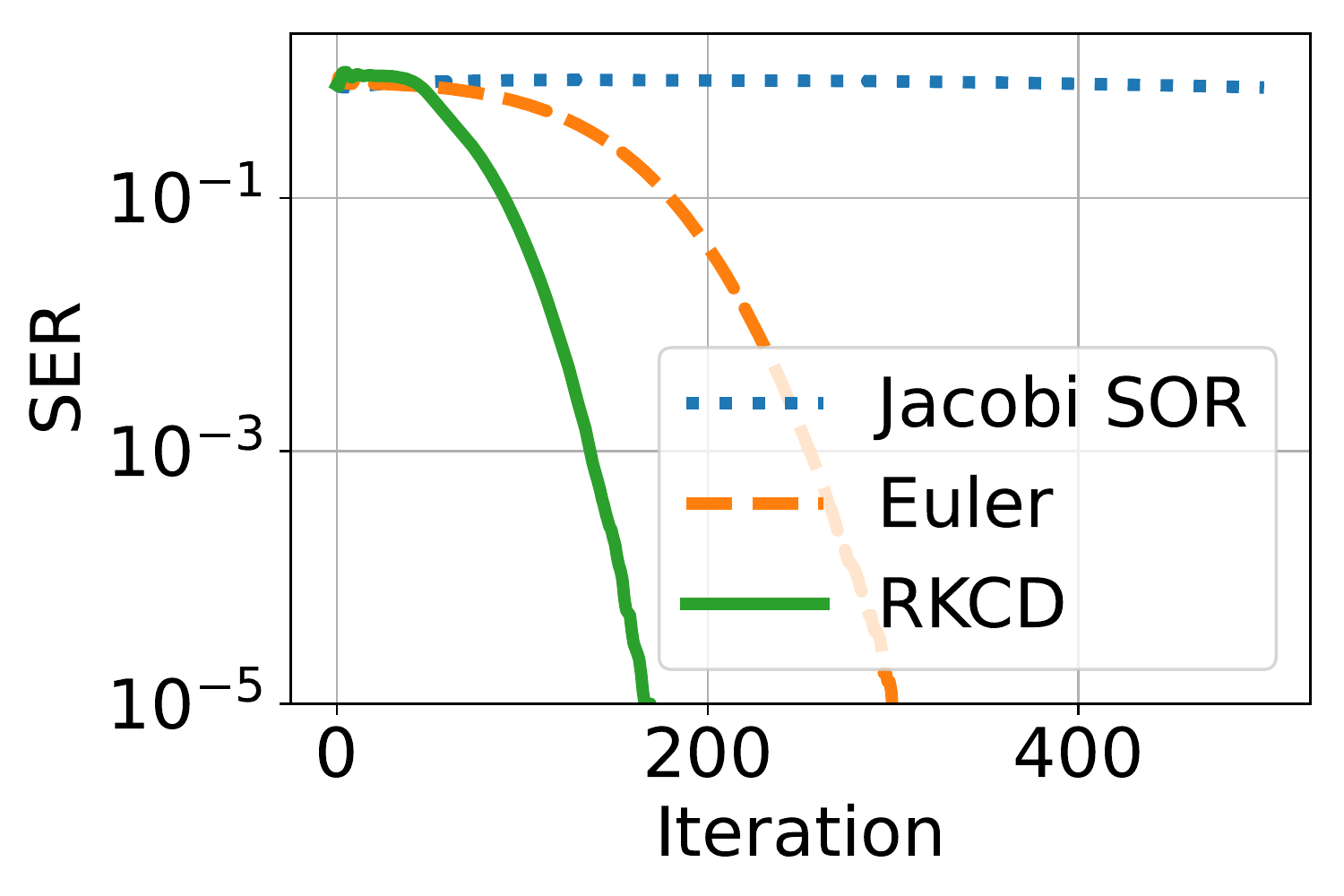}
      \end{minipage} & 
      \begin{minipage}[b]{0.45\hsize}
        \centering
        \includegraphics[width=\columnwidth]{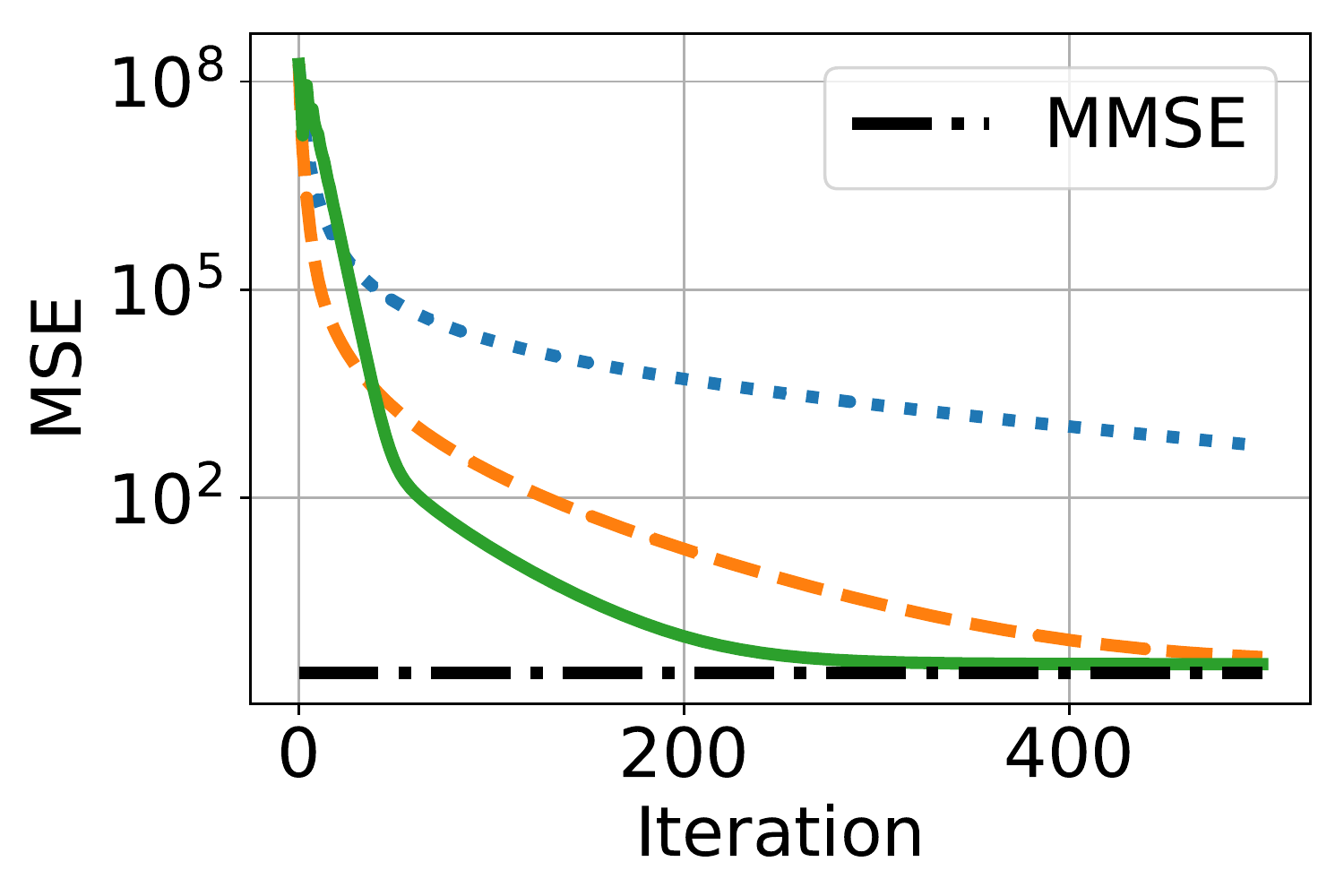}
      \end{minipage}
    \end{tabular}
    \caption{Comparison of discrete-time MMSE estimation methods, $(n,m,\sigma^2)=(400,500,0.1)$, 16QAM.}
    \label{fig:rkc2}
\end{figure}
Figures~\ref{fig:rkc} and \ref{fig:rkc2} show the SER and MSE performance of 
the Euler method-based algorithm \eqref{eq:euler}, 
RKCD method-based algorithm (Algorithm~\ref{alg:rkc}), and conventional Jacobi SOR algorithm.
The system parameters and damping constants were set to $(n,m,\sigma^2,\eta)=(60,80,0.1,0.1)$, $\epsilon=2.3$ in Fig.~\ref{fig:rkc} 
and $(n,m,\sigma^2,\eta)=(400,500,0.1,0.1)$, $\epsilon=5$ in Fig.~\ref{fig:rkc2}.
The step-size parameter $\delta$ in the Euler method was set to $\delta=0.001$ in both cases.
For the RKCD method, the parameters $s$ and $h$ were set as in \eqref{eq:sandh}, respectively, and 
we employed accurate eigenvalues $\lambda_1$ and $\lambda_n$ as the lower bound $\ell$ and upper bound $L$, respectively.
The arithmetic MSE was calculated for $10000$ and $1000$ times generation of channel matrices $\bm{H}$ and signals $\bm{y}, \bm{x}$ for Fig.~\ref{fig:rkc} and \ref{fig:rkc2}, respectively, 
using Monte Carlo simulation.
We used 16QAM signals.
From Figs.~\ref{fig:rkc} and \ref{fig:rkc2}, 
the slopes of the MSE curves of the Jacobi SOR and Euler methods are large in a small number of iterations, 
but then decrease immediately.
On the other hand, the RKCD method maintained a large slope. 
It leads to faster convergence and a large performance gap compared with other methods.
The results show that the RKCD method-based algorithm achieves better performance than other MMSE detection algorithms.

\begin{figure}[tb]
    \begin{tabular}{cc}
      \begin{minipage}[b]{0.45\hsize}
        \centering
        \includegraphics[width=\columnwidth]{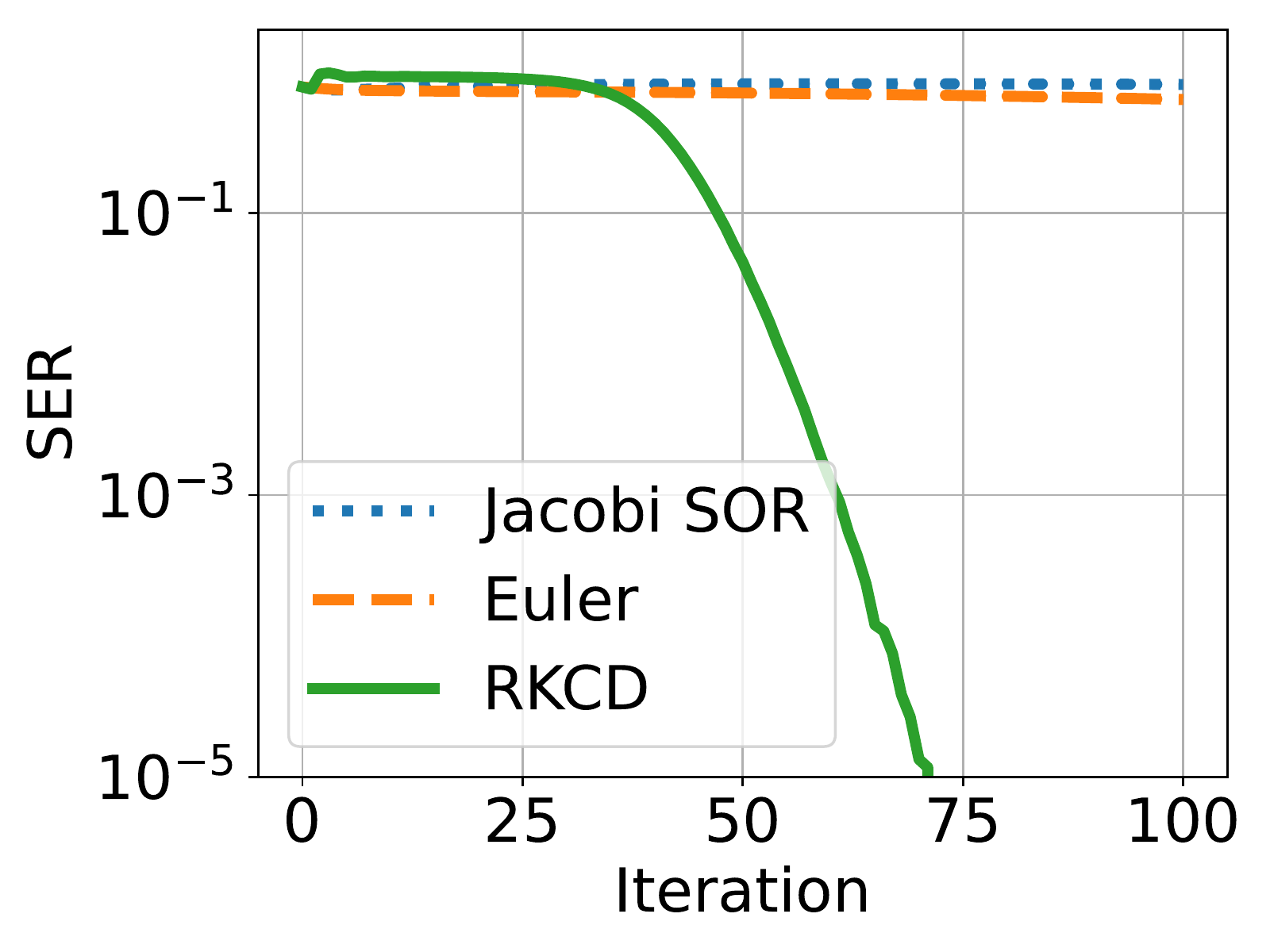}
      \end{minipage} & 
      \begin{minipage}[b]{0.45\hsize}
        \centering
        \includegraphics[width=\columnwidth]{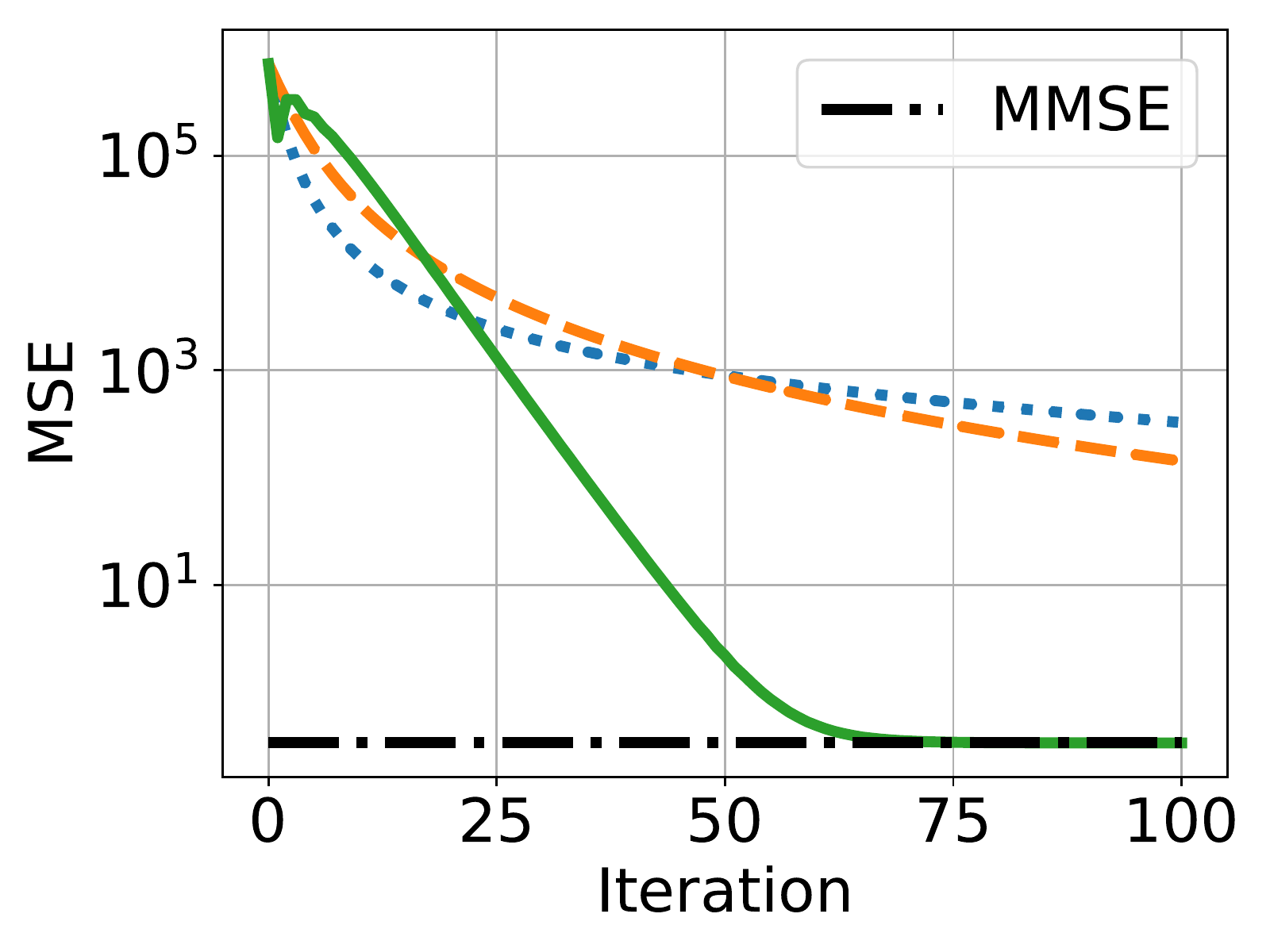}
      \end{minipage}
    \end{tabular}
    \caption{Comparison of discrete-time MMSE estimation methods using Kronecker channel model, $(n,m,\sigma^2)=(60,80,0.1)$, 16QAM.}
    \label{fig:rkckron}
\end{figure}
\begin{figure}[tb]
    \begin{tabular}{cc}
      \begin{minipage}[b]{0.45\hsize}
        \centering
        \includegraphics[width=\columnwidth]{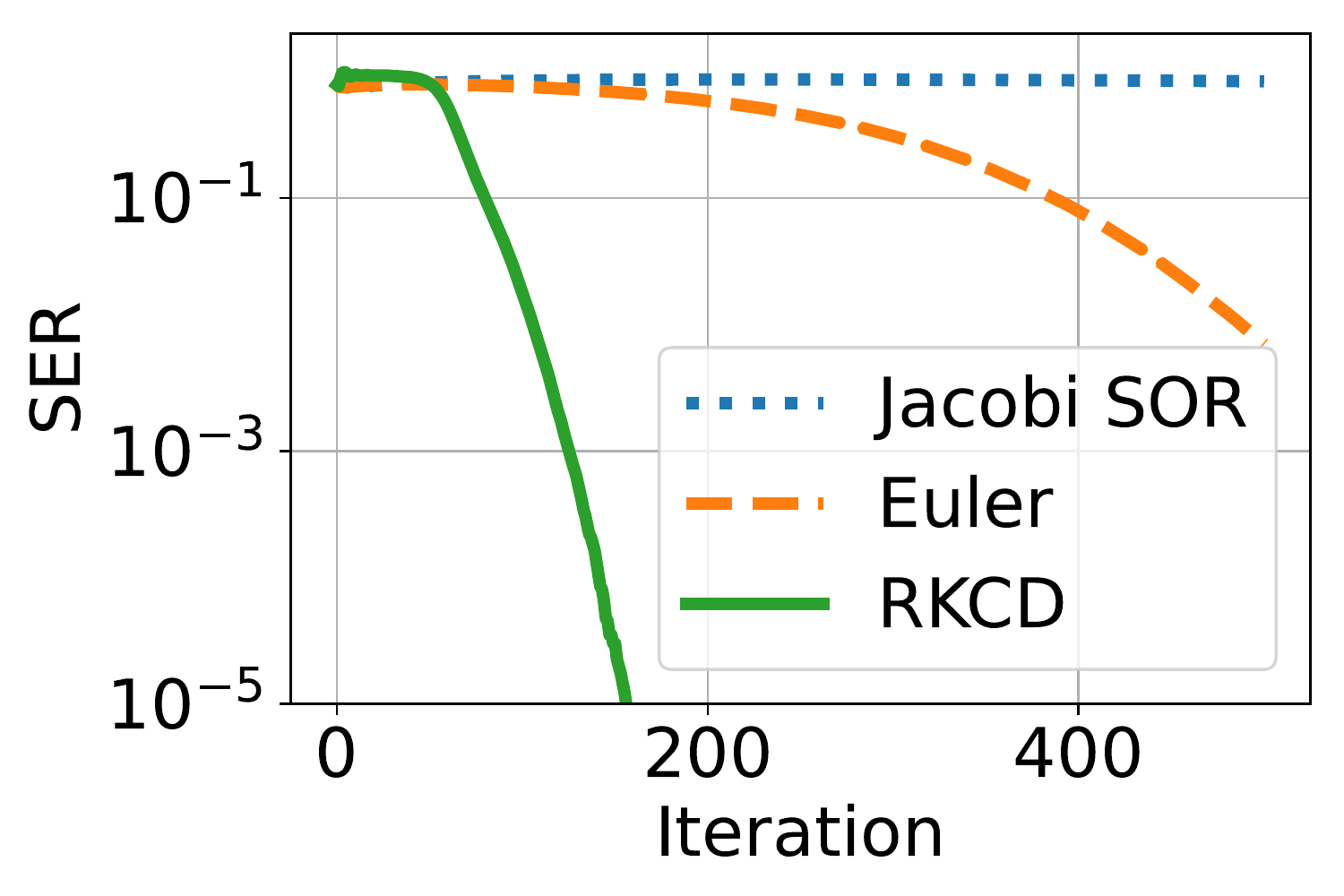}
      \end{minipage} & 
      \begin{minipage}[b]{0.45\hsize}
        \centering
        \includegraphics[width=\columnwidth]{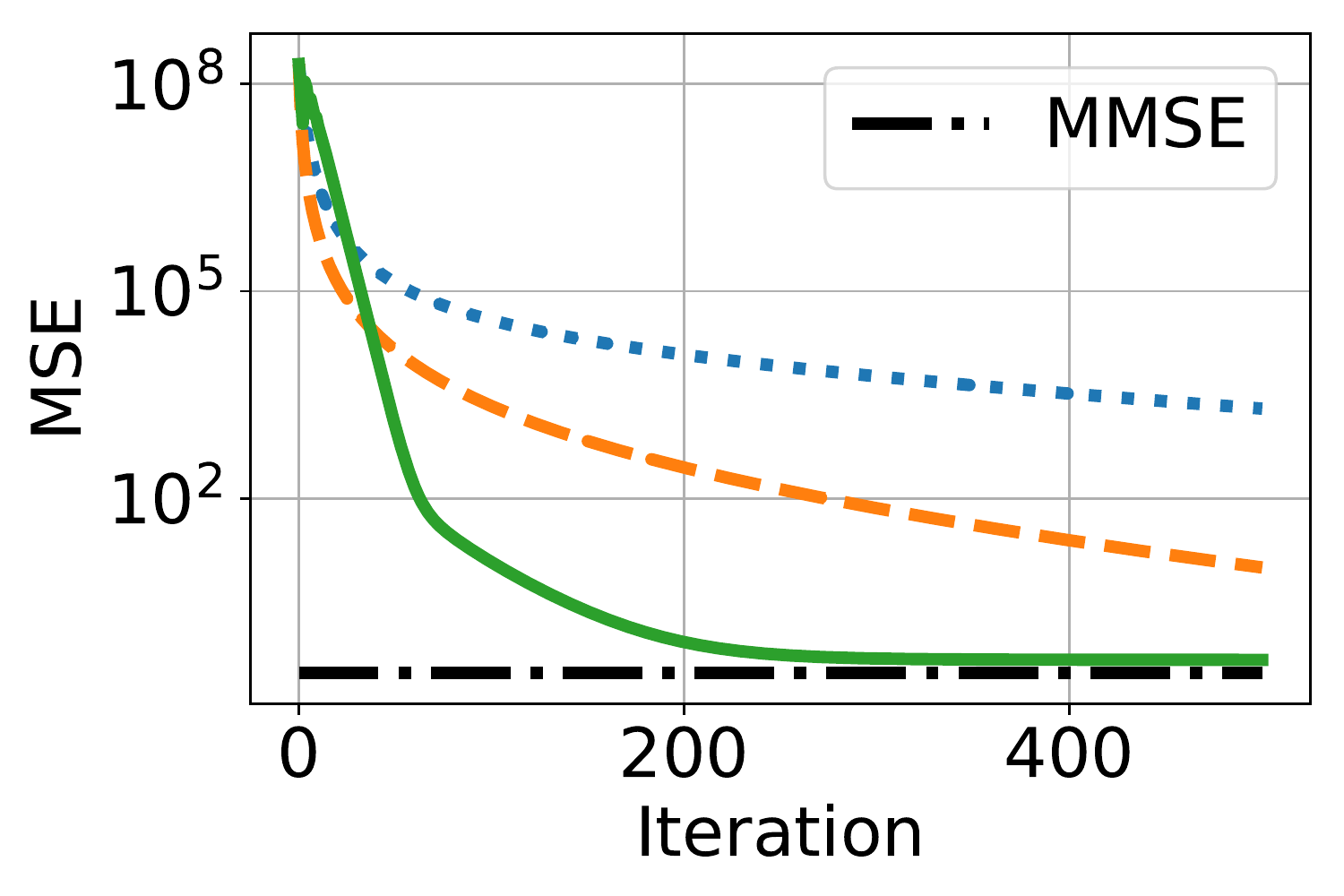}
      \end{minipage}
    \end{tabular}
    \caption{Comparison of discrete-time MMSE estimation methods using Kronecker channel model, $(n,m,\sigma^2)=(400,500,0.1)$, 16QAM.}
    \label{fig:rkc2kron}
\end{figure}
Convergence performance of another channel model is shown in Figs.~\ref{fig:rkckron} and \ref{fig:rkc2kron}.
We used the Kronecker channel model with exponential correlation \cite{Loyka}.
In the model, the channel matrix $\bm{H}$ was generated by 
\begin{equation}
    \bm{H}=\bm{R}_\mathrm{R}^{1/2}\bm{G}\bm{R}_\mathrm{T}^{1/2},
\end{equation}
where $\bm{R}_\mathrm{R}\in\mathbb{R}^{m\times m}$ and $\bm{R}_\mathrm{T}\in\mathbb{R}^{n\times n}$ are spatial correlation matrices 
with correlation coefficient $\rho=0.2$, 
and where the matrix $\bm{G}\in\mathbb{C}^{m\times n}$ was generated so that each element followed $\mathcal{CN}(0,1)$.
The same system parameters were used as in Figs.~\ref{fig:rkc} and \ref{fig:rkc2}.
The damping constant $\epsilon$ for the RKCD method-based algorithm and step-size parameter $\delta$ for the Euler method 
were set to $(\epsilon,\delta)=(1.5,0.001)$ for Fig.~\ref{fig:rkckron} 
and $(\epsilon,\delta)=(5,0.0005)$ for Fig.~\ref{fig:rkc2kron}, respectively.
The other settings were the same as those in Figs.~\ref{fig:rkc} and \ref{fig:rkc2}.
The RKCD method-based algorithm also shows a similar performance tendency to Figs.~\ref{fig:rkc} and \ref{fig:rkc2} 
in another channel model.

As the numerical results show, 
it is possible to create algorithms 
that exhibit performance not achievable using conventional discrete-time algorithms 
by constructing an ODE to solve the problem of interest 
and discretizing the ODE with ingenious step-size parameters.

\subsubsection{Analysis of Discrete-time Algorithm}
\begin{figure}[tbp]
    \centerline{\includegraphics[width=\columnwidth]{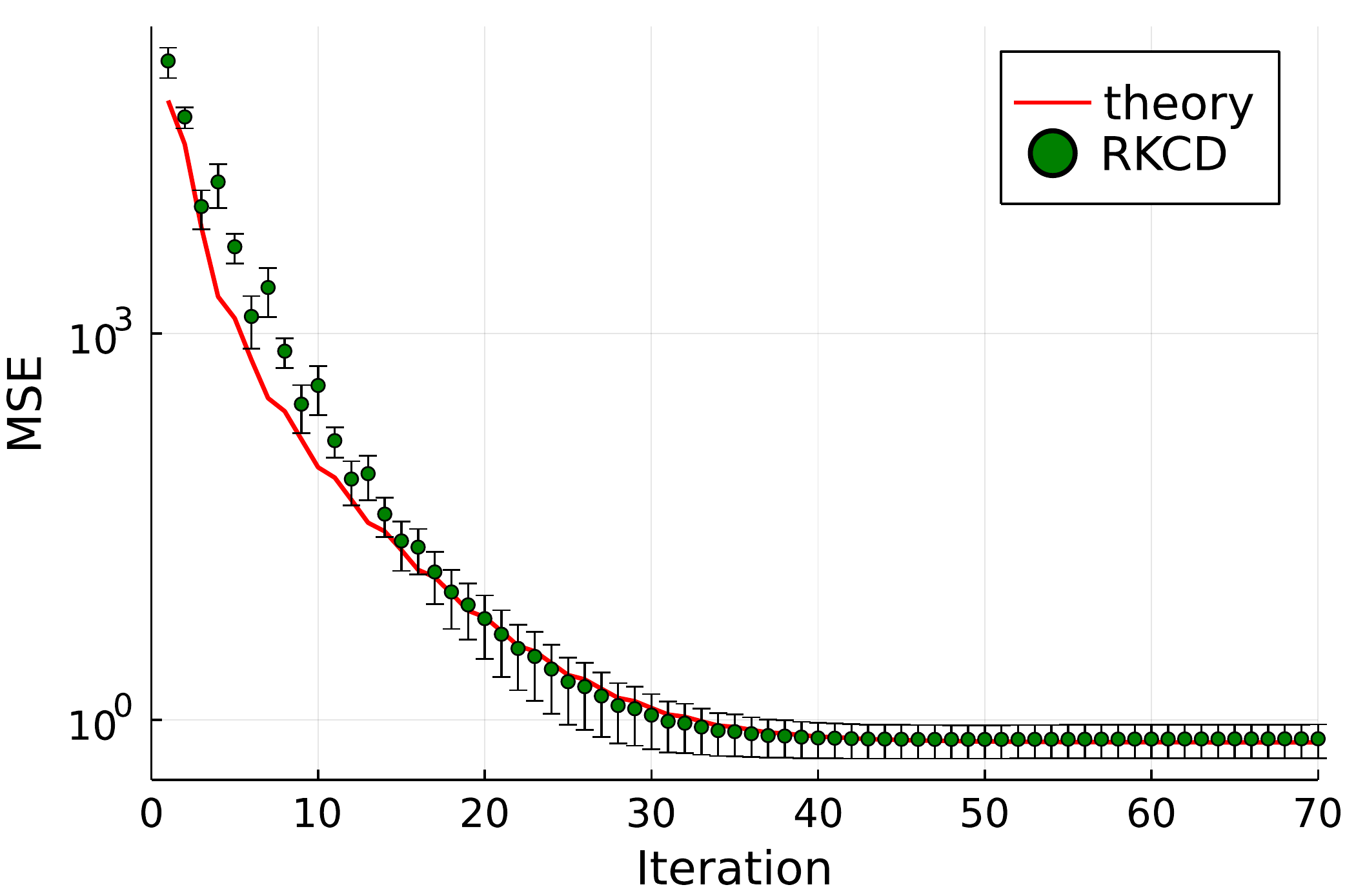}}
    \caption{MSE obtained by MSE formula and arithmetic MSE of RKCD, $(n,m,\sigma^2,\kappa)=(20,50,1,13.93)$, QPSK.}
    \label{fig:theoryandrkc}
\end{figure}

We present a numerical example of MSE analysis of the discrete-time algorithm.
Figure~\ref{fig:theoryandrkc} shows the MSE values obtained from the MSE formula \eqref{eq:mse} and 
arithmetic MSE values obtained using the RKCD method-based estimation method.
The system parameters were set to $(n,m,\sigma^2)=(20,50,1)$ and 
the regularization parameter was $\eta=1$.
The condition number of the Gram matrix was $\kappa=13.93$ in this case.
The parameter $s$ for the RKCD method was determined by \eqref{eq:sandh} and 
$h$ was set to $h=0.03185$.
We used QPSK signals.
We performed 100 trials to calculate the arithmetic MSE values, which are displayed as markers in the figure. 
The standard deviation of the squared error is shown as error bars.
From Fig.~\ref{fig:theoryandrkc}, the arithmetic MSE values are close to the theoretical values.
In other words, the performance behavior of the algorithm can be well described using the MSE formula.

\section{Conclusions}
\label{sec:conc}
We have explored continuous-time MMSE signal detection methods for MIMO systems assuming the implementation by analog optical devices 
as a potential solution to computational load issues 
in future wireless communication systems.
We described the continuous-time estimation as an ODE and proposed the ODE-MMSE method.
The analytical formula for the MSE was derived 
by using the eigenvalue decomposition of the Gram matrix of the channel matrix.
We also extended the ODE-MMSE method by introducing the time-dependent parameter 
$\eta(t)$ and proposed the tODE-MMSE method. 
The MSE formula was derived also for the tODE-MMSE method.

Signal processing with analog optical devices has excellent potential for improving computational efficiency 
and overcoming many of the limitations of traditional digital signal processing.
However, optical devices are known to include system noises 
and they may cause detection errors.
Stochastic differential equations will need to be employed 
to analyze the behavior of the noisy system in detail.
Theoretical analysis of the continuous-time signal detection method 
presented in this paper 
is expected to render analog computation a more realistic technology for next-generation communication systems.
The findings of this study can inspire further research in the development of analog devices for signal processing.

This paper also presented the byproducts of considering a continuous-time system 
on discrete-time detection algorithms.
{{Although any numerical method can be applied for discretizing ODE, 
the proposed RKCD method-based algorithm 
achieves better convergence performance without any additional computational costs.}} 
The MSE of the discrete-time algorithm has become analytically traceable using the MSE formula for the ODE-MMSE method.
This advantage reveals that 
continuous-time signal processing and its analysis have the potential to be a new construction methodology for discrete-time algorithms.
The approach proposed in this study can lead to the development of more efficient and accurate discrete-time signal processing algorithms 
and a deeper understanding of the fundamental principles of signal processing, 
which can have a wide range of practical applications.

{\appendices
\section{Derivation of Theorem~\ref{theo:analyticalmse}}
\label{sec:appa}
The matrix $\bm{Q}(t)$ is given by 
\eqref{eq:Qt}.
By applying eigenvalue decomposition to $(\bm{H}^\mathrm{H}\bm{H}+\eta\bm{I})^{-1}$ and 
$\exp{(-(\bm{H}^\mathrm{H}\bm{H}+\eta\bm{I})t)}$, 
the matrix $\bm{Q}(t)$ can be expanded as 
$\bm{Q}(t) = \bm{U}\mathrm{diag}[q(\lambda_1)/(\lambda_1+\eta),\ldots, q(\lambda_n)/(\lambda_n+\eta)]\bm{U}^\mathrm{H}\bm{H}$, 
where $q(\lambda_i):=e^{-(\lambda_i+\eta)t}(\lambda_i+\eta-1)+1$.
The term \eqref{eq:tmp1} can be calculated by using this and the relation $\mathrm{Tr}[\bm{UXU}^\mathrm{H}]=\mathrm{Tr}[\bm{X}]$ for a matrix $\bm{X}$ as 
\begin{align}
	&\mathrm{Tr}[\bm{Q}(t)^\mathrm{H}\bm{Q}(t)] \nonumber \\
    &= \mathrm{Tr}\left[\mathrm{diag}\left[\left(\frac{q(\lambda_1)}{\lambda_1+\eta}\right)^2\lambda_1,\ldots, \left(\frac{q(\lambda_n)}{\lambda_n+\eta}\right)^2\lambda_n\right]\right] \\
    &= \sum_{i=1}^n \frac{\lambda_i\left(q(\lambda_i)\right)^2}{(\lambda_i+\eta)^2}.
\end{align}
The term \eqref{eq:tmp2} can be also expanded as follows: 
\begin{equation}
    \mathrm{Tr}\left[(\bm{Q}(t)\bm{H}-\bm{I})^\mathrm{H}(\bm{Q}(t)\bm{H}-\bm{I})\right] 
    = \sum_{i=1}^n \frac{\left(\lambda_iq(\lambda_i)-(\lambda_i+\eta)\right)^2}{(\lambda_i+\eta)^2}.
\end{equation}
By substituting them into \eqref{eq:tmpmse}, 
\begin{align}
	&\mathrm{MSE}(t) \nonumber \\
    &= \sum_{i=1}^n \frac{\left(\lambda_iq(\lambda_i)-(\lambda_i+\eta)\right)^2}{(\lambda_i+\eta)^2} + \sigma^2 \sum_{i=1}^n \frac{\lambda_i\left(q(\lambda_i)\right)^2}{(\lambda_i+\eta)^2} \\
    &= \sum_{i=1}^n \frac{\lambda_i(\lambda_i+\eta-1)^2(\lambda_i+\sigma^2)e^{-2(\lambda_i+\eta)t}}{(\lambda_i+\eta)^2} \nonumber \\
    &\ - \sum_{i=1}^n \frac{2\lambda_i(\lambda_i+\eta-1)(\eta-\sigma^2)e^{-(\lambda_i+\eta)t}}{(\lambda_i+\eta)^2} \nonumber \\
    &\ + \sum_{i=1}^n \frac{\eta^2+\sigma^2\lambda_i}{(\lambda_i+\eta)^2}.
\end{align}

\section{Derivation of Lemma~1}
\label{sec:appb}
The MSE of MMSE estimation is defined by 
$\mathrm{MSE}_\mathrm{mmse} =\mathbb{E}[\|\hat{\bm{x}}-\bm{s}\|^2]$.
By substituting \eqref{eq:mmse} into the definition, 
\begin{align}
    \mathrm{MSE}_\mathrm{mmse} 
    &=\mathrm{Tr}\Bigl[\left(\left(\bm{H}^\mathrm{H}\bm{H}+\sigma^2\bm{I}\right)^{-1}\bm{H}^\mathrm{H}\bm{H}-\bm{I}\right)^\mathrm{H} \nonumber \\
    & \quad \cdot\left(\left(\bm{H}^\mathrm{H}\bm{H}+\sigma^2\bm{I}\right)^{-1}\bm{H}^\mathrm{H}\bm{H}-\bm{I}\right)\Bigr] \nonumber \\
    & \ + \sigma^2\mathrm{Tr}\Bigl[\left(\left(\bm{H}^\mathrm{H}\bm{H}+\sigma^2\bm{I}\right)^{-1}\bm{H}^\mathrm{H}\right)^\mathrm{H} \nonumber \\
    & \quad \cdot\left(\left(\bm{H}^\mathrm{H}\bm{H}+\sigma^2\bm{I}\right)^{-1}\bm{H}^\mathrm{H}\right)\Bigr], 
\end{align}
where the second moments are $\mathbb{E}[\bm{ss}^\mathrm{H}]=\bm{I}$ and $\mathbb{E}[\bm{ww}^\mathrm{H}]=\sigma^2\bm{I}$.
By applying eigenvalue decomposition to $\left(\bm{H}^\mathrm{H}\bm{H}+\sigma^2\bm{I}\right)^{-1}$ 
and $\bm{H}^\mathrm{H}\bm{H}$, 
\begin{align}
    &\mathrm{MSE}_\mathrm{mmse} \nonumber\\
    &= \sum_{i=1}^n \left(\frac{\lambda_i}{\lambda_i+\sigma^2}-1\right)^2 
    + \sigma^2\sum_{i=1}^n \left(\frac{\lambda_i}{(\lambda_i+\sigma^2)^2}\right) \\
    &= \sum_{i=1}^n \frac{\sigma^2}{\lambda_i+\sigma^2}.
\end{align}

\section{Derivation of Theorem~2}
\label{sec:appc}
The matrix integral in \eqref{eq:solutiontime} can be decomposed as 
\begin{align}
    &\int_0^t e^{\bm{H}^\mathrm{H}\bm{H}u+\xi(u) \bm{I}}du \nonumber \\
    &= \bm{U} \mathrm{diag}\left[\int_0^t e^{\lambda_1u+\xi(u) }du, \ldots,\int_0^t e^{\lambda_nu+\xi(u) }du\right]\bm{U}^\mathrm{H}.
\end{align}
By using this, the matrix included in \eqref{eq:solutiontime} is also decomposed as 
\begin{align}
    &\exp{\left(-\bm{H}^\mathrm{H}\bm{H}t-\xi(t) \bm{I}\right)}\left(\bm{I}+ \int_0^t e^{\bm{H}^\mathrm{H}\bm{H}u+\xi(u) \bm{I}}du\right) \nonumber \\
    &= \bm{U}\Bigl(\mathrm{diag}\Bigl[e^{-(\lambda_1 t+\xi(t))}\left(1+\int_0^t e^{\lambda_1 u+\xi(u)}du\right), \ldots, \nonumber \\
    & \quad e^{-(\lambda_n t+\xi(t))}\left(1+\int_0^t e^{\lambda_n u+\xi(u)}du\right)\Bigr]\Bigr)\bm{U}^\mathrm{H}.
\end{align}
Applying this to the definition of MSE leads to the analytical formula in the same way as Theorem~1.

\section{Numerical Method for Emulating Continuous-time Behavior}
\label{sec:appd}
We employed the well-known Euler method for {{emulating}} the continuous-time behavior of $\bm{x}(t)$. 
The Euler method discretizes time window $[0,t_{\max}]$ with $t_{\max}/\delta$ bins, where $\delta$ is step-size and to be set to a sufficiently small value.
The estimate at time $t_k=\delta k \ (k=1,2,\ldots,t_{\max}/\delta)$ 
is given by \eqref{eq:euler}.

\begin{figure}[tbp]
    \centerline{\includegraphics[width=\columnwidth]{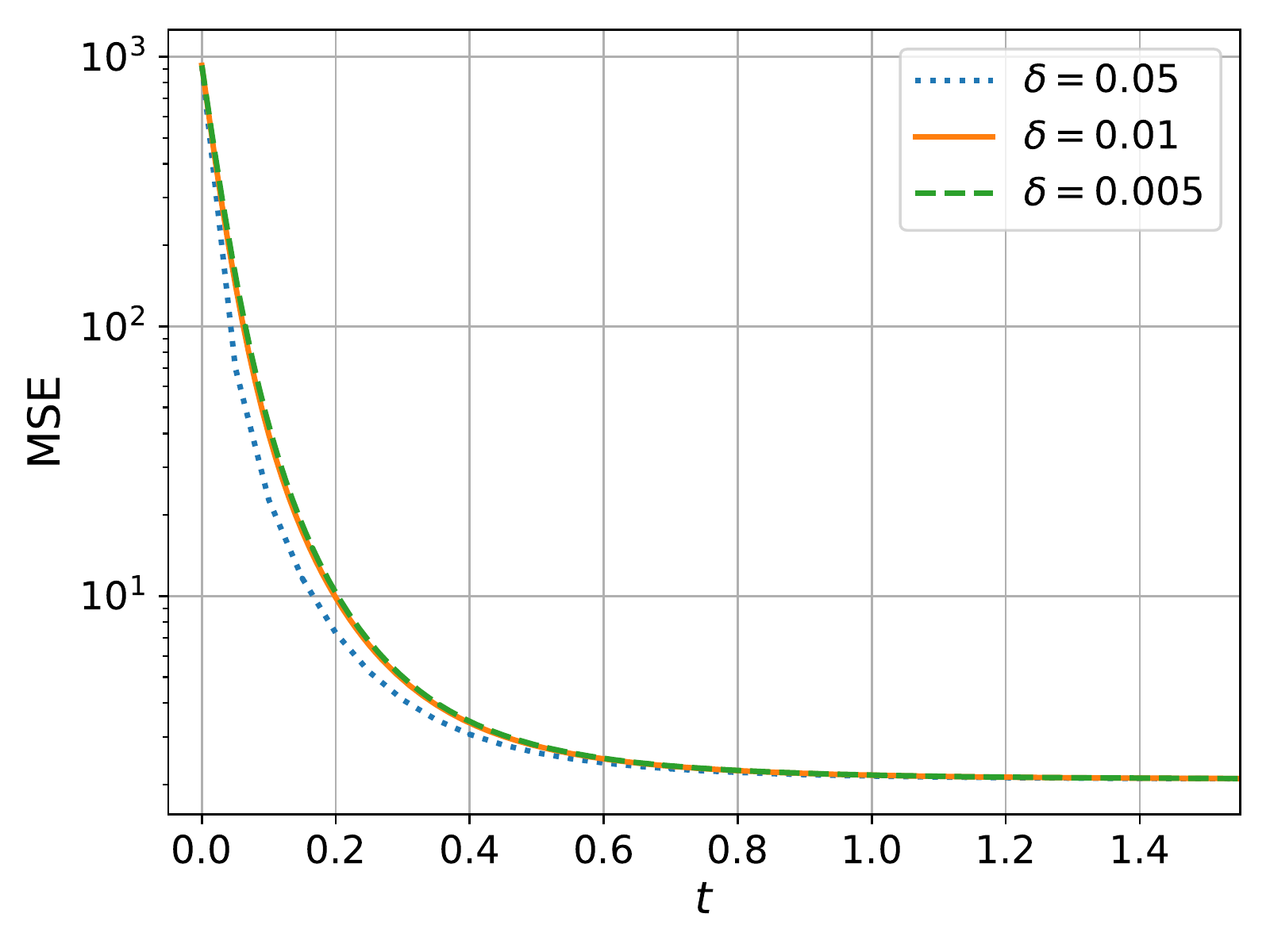}}
    \caption{Evaluation of accuracy of numerical method depending on step-size values, $(n,m,\sigma^2,\eta,t_{\max},\kappa)=(8,8,1,0.5,3)$, QPSK.}
    \label{fig:delta}
\end{figure}
As a preliminary experiment, we have verified a sufficient value for the step size $\delta$ 
to accurately simulate the ODE \eqref{eq:ode} 
because too large $\delta$ may lead to inaccurate behavior.
The system parameters were set to $(n,m,\sigma^2,\eta)=(8,8,1,0.5)$.
We set $t_{\max}=3$ and $\delta=0.05, 0.01$, and $0.005$.
We generated a single instance of the channel matrix $\bm{H}$, 
where each element follows an independent and identically distributed $\mathcal{CN}(0,1)$.
The condition number of the Gram matrix $\bm{H}^\mathrm{H}\bm{H}$ was $\kappa=106.82$.
For the Monte Carlo simulation, 
QPSK signal $\bm{s}$, noise $\bm{w}$, and the corresponding received signal $\bm{y}$ 
were generated $1000$ times, 
and then the arithmetic MSE, 
the estimate of MSE calculated from arithmetic mean of squared errors, was computed 
with the matrix $\bm{H}$ fixed.
From Fig.~\ref{fig:delta}, 
we confirmed that there is no visible difference between the MSE curves for $\delta=0.01$ and $\delta=0.005$, 
so that we set $\delta$ to be smaller than 0.01 in the simulations.

\section{Comparison of RKCD Method-based MMSE with Other Detection Methods}
\label{sec:appe}
In this section, 
we compare the detection performance of the RKCD method-based MMSE detection (Algorithm~\ref{alg:rkc}) 
with that of recent detection methods, OAMP-based detection \cite{oamp} {{including matrix inverse calculation}} 
and OAMP-Net2 \cite{He} {{requiring learning}}, 
{{to confirm the competitiveness against state-of-the-art algorithms}}.
As noted in Sect. \ref{sec:discreteintro}, 
these recent algorithms generally show better performance than MMSE 
at the cost of the matrix inverse calculation per iteration.
Figure~\ref{fig:discrete} shows the SER performance of the detection methods 
when system parameters and the damping constant were set to $(n,m,\sigma^2,\eta)=(16,32,0.5,0.5)$ and $s=20$.
We used QPSK signals.
To align with the implementation requirements of the conventional methods, 
the initial value of the estimate was set to $\bm{x}^{[0]}=\bm{0}$, 
and the channel matrix was generated so that each element followed $\mathcal{CN}(0,1/m)$.
The RKCD method-based MMSE detection, 
{{which includes neither matrix inverse calculation nor learning,}} 
shows comparable SER with a comparable number of iterations 
to OAMP and OAMP-Net2 under these system settings.
In other words, 
this method may achieve comparable performance with low computational complexity 
to that of recent detection methods, 
depending on channel environments.
\begin{figure}[tbp]
    \centerline{\includegraphics[width=\columnwidth]{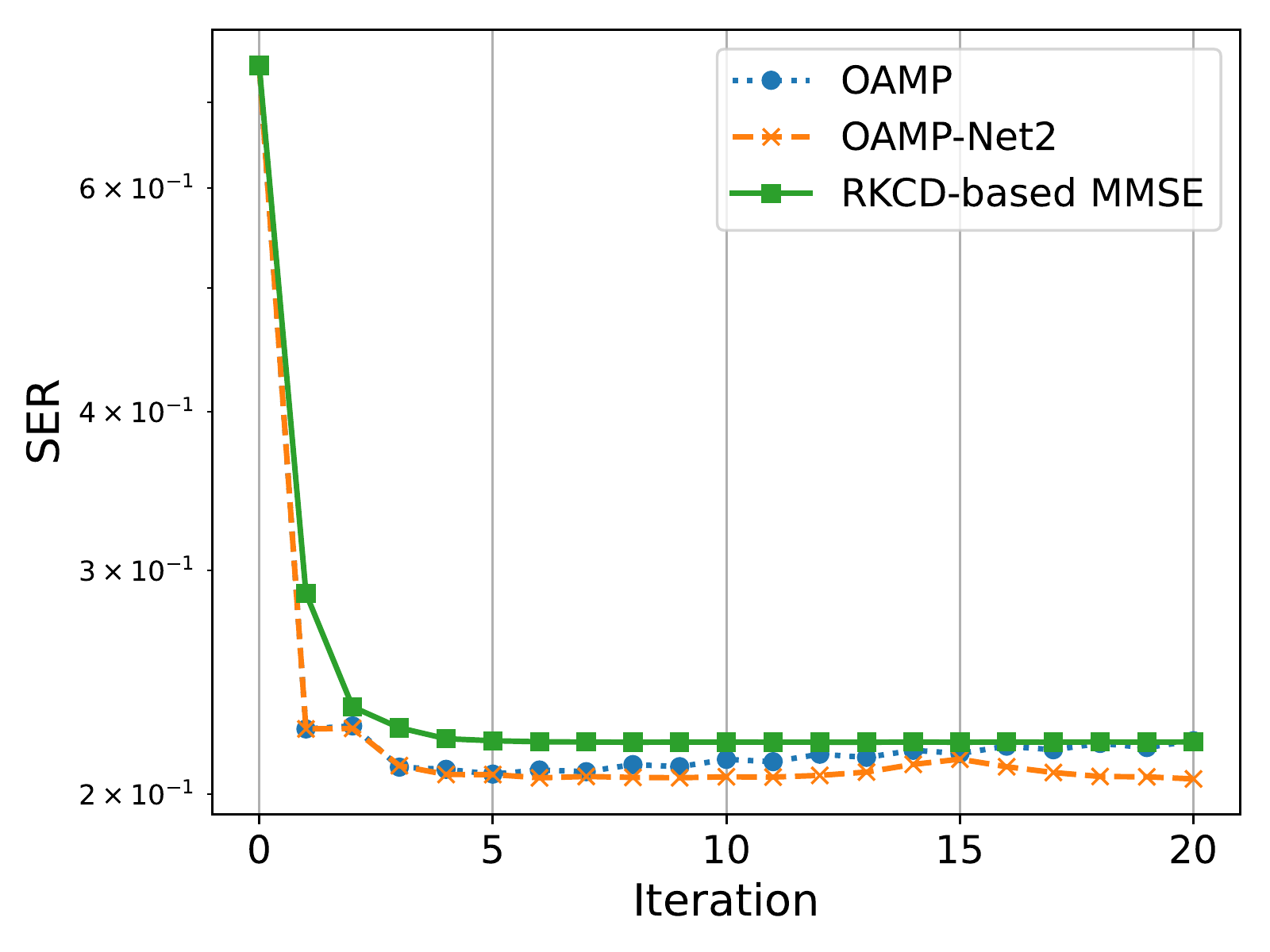}}
    \caption{Comparison with recent detection methods, QPSK.}
    \label{fig:discrete}
\end{figure}
}

\section*{Acknowledgments}
This work was supported by 
JSPS KAKENHI Grant-in-Aid for Young Scientists Grant Number JP23K13334 (to A. Nakai-Kasai) 
and Scientific Research(A) Grant Number JP22H00514 (to T. Wadayama).

 




\vspace{11pt}
\begin{IEEEbiography}[{\includegraphics[width=1in,height=1.25in,clip,keepaspectratio]{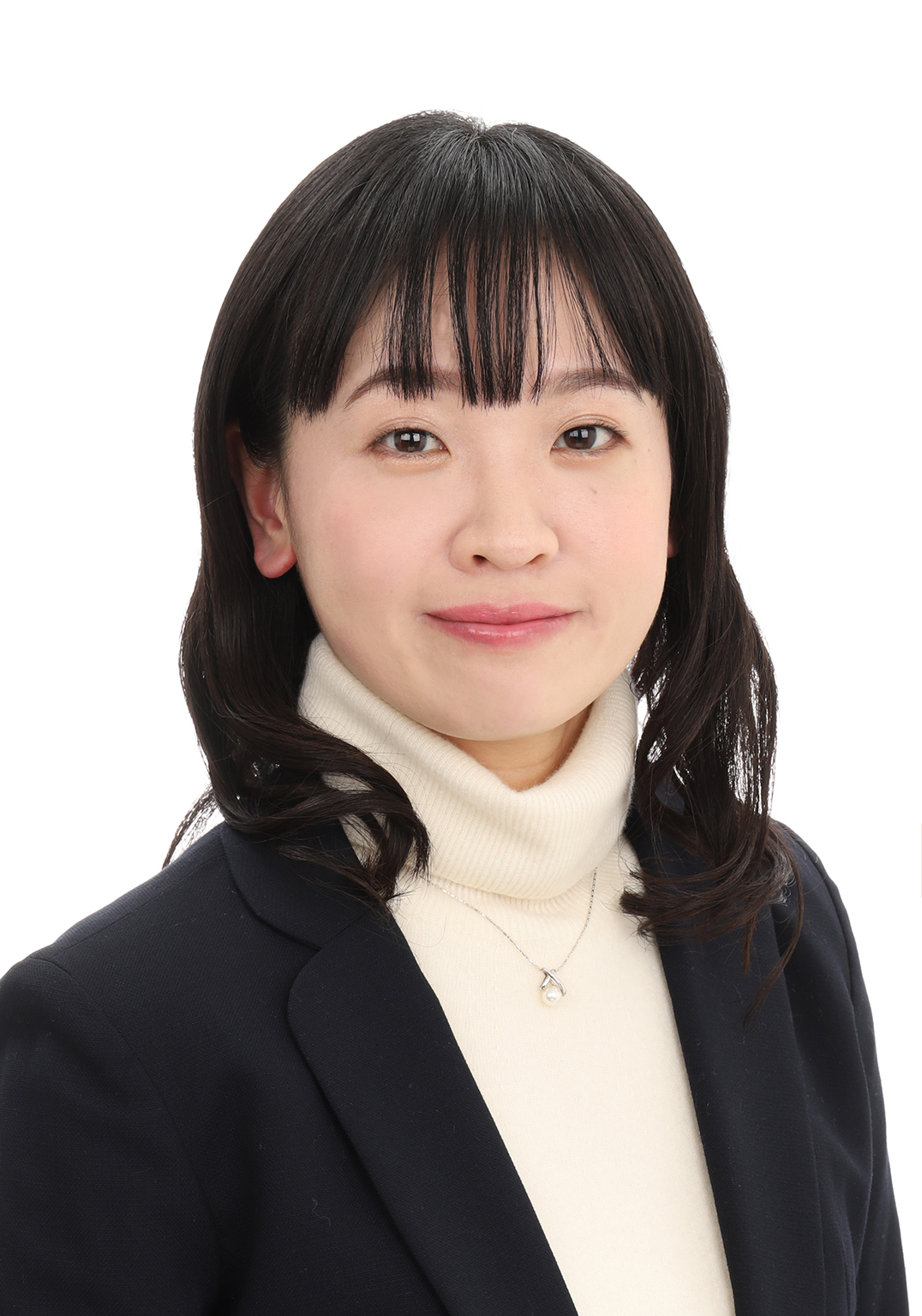}}]{Ayano Nakai-Kasai}
    received the bachelor’s degree 
    in engineering, the master’s degree in informatics,
    and Ph.D. degree in informatics from Kyoto University, Kyoto, Japan, in 2016, 2018, and 2021,
    respectively. She is currently an Assistant Professor
    at Graduate School of Engineering, Nagoya Institute of Technology. Her research interests include
    signal processing, wireless communication, and machine learning. She received the Young Researchers’
    Award from the Institute of Electronics, Information 
    and Communication Engineers in 2018 and APSIPA 
    ASC 2019 Best Special Session Paper Nomination Award. She is a member
    of IEEE and IEICE.
\end{IEEEbiography}
\begin{IEEEbiography}[{\includegraphics[width=1in,height=1.25in,clip,keepaspectratio]{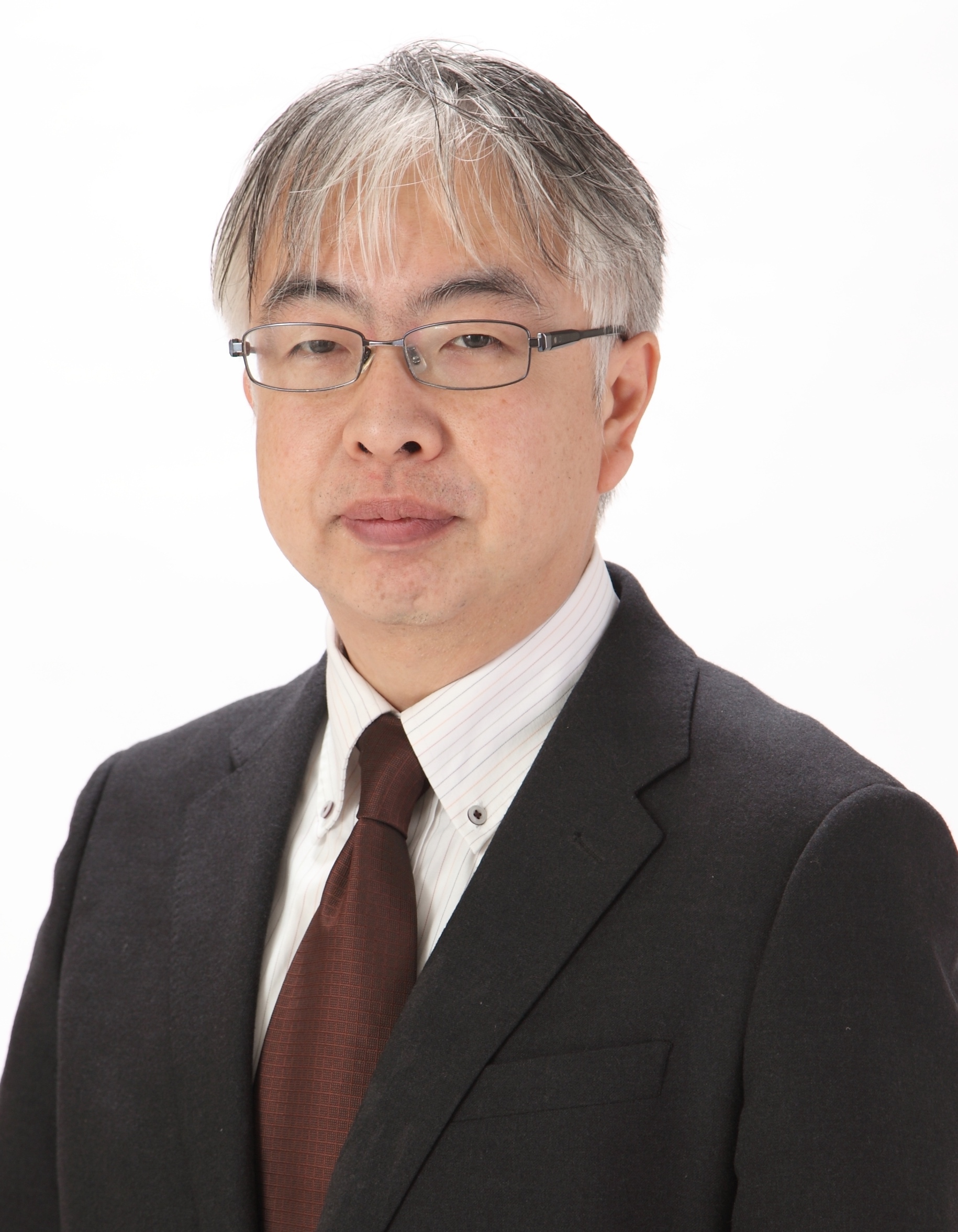}}]{Tadashi Wadayama}
    (M'96) was born in Kyoto, Japan, on May 9,1968.
    He received the B.E., the M.E., and the D.E. degrees from Kyoto Institute of Technology in 1991, 1993 and 1997, respectively.
    On 1995, he started to work with Faculty of Computer Science and System Engineering, Okayama Prefectural University as a research associate.
    From April 1999 to March 2000, he stayed in Institute of Experimental Mathematics, Essen University (Germany) as a visiting researcher.
    On 2004, he moved to Nagoya Institute of Technology as an associate professor. Since 2010, he has been a full professor of Nagoya Institute of Technology.
    His research interests are in coding theory, information theory, and signal processing for wireless communications.
    He is a member of IEEE and a senior member of IEICE.
\end{IEEEbiography}

\vfill

\end{document}